\documentclass[a4paper,useAMS,usenatbib]{mn2e}

\usepackage{times}

\usepackage{setspace}

\usepackage{comment}

\usepackage{epsfig}

\usepackage{natbib}

\def \mnras {{MNRAS}}

\def \apj {{ApJ}}

\def \apjs {{ApJS}}

\def \apss {{Ap\&SS}}

\def \aap {{A\&A}}

\def \aj {{AJ}}

\def \apjl {{ApJL}}

\def \araa {{ARA\&A}}

\title[Secular Evolution Rate for Spiral Galaxies]{An Observational Estimate for the Mean Secular Evolution Rate in Spiral Galaxies}

\author[Foyle,  Rix, \& Zibetti]{Kelly Foyle$^{1}$\thanks{E-mail: foyle@mpia.de}, Hans-Walter Rix$^{1}$ \& Stefano Zibetti$^{1}$\\  $^{1}$Max-Planck-Institut f\"ur Astronomie, K\"onigstuhl 17, Heidelberg, Germany}

\date{accepted}

\pagerange{\pageref{firstpage}--\pageref{lastpage}}\pubyear{2009}

\begin{document}

\label{firstpage}


\maketitle

\begin{abstract} 
We have observationally quantified the effect of gravitational torques
on stars in disk galaxies due to the stellar distribution itself and
explored whether these torques are efficient at transporting angular
momentum within a Hubble Time.  We derive instantaneous torque maps for
a sample of 24 spiral galaxies, based on stellar mass maps that were
derived using the pixel-by-pixel mass-to-light estimator by Zibetti, Rix
and Charlot.  In conjunction with an estimate of the rotation velocity,
the mass maps allow us to determine the torque-induced instantaneous
angular momentum flow across different radii, resulting from the overall
stellar distributions for each galaxy in the sample.  By stacking the
sample, which effectively replaces a time average by an ensemble
average, we find that the torques due to the stellar disk act to
transport angular momentum outward over much of the disk (within 3 disk
scale lengths).  The strength of the ensemble-averaged gravitational
torques within one disk scale length have a timescale of  ~ 4 Gyr for
angular momentum redistribution.

The individual torque profiles show that only a third of our sample
exhibit torques strong enough to redistribute angular momentum within a
Hubble Time, mostly those with strong bars.  However, advective angular
momentum transport is another source of angular momentum redistribution,
especially in the presence of long-lived spiral arms, but is not
accessible to direct observations.  The torque-driven angular momentum
redistribution is thus observed to be effective, either in one third of
disk galaxies or in most disk galaxies one third of the time and should
lead to either changes in the mass density profile or the orbital shapes.

We use a set of self-consistent disk, bulge and halo simulated isolated
disk galaxies with realistic cold gas fractions to verify that the
torques exerted by the stellar distribution, such as spiral arms or a
bar, exceed those of the gas and halo, as assumed in the analysis of the
observations.

This study is the first to observationally determine the strength of
 torque-driven angular momentum flow of stars for a sample of
spiral galaxies,  providing an important empirical constraint on
secular evolution.

\end{abstract}




\section{Introduction}

Three primary aspects control the observable features of present-day galaxies: (1) the particular density fluctuations from which they originated, (2) successive mergers and various interactions with the environment and (3) the (internal) secular evolution.  While it is clear that the first and second point play a decisive role in shaping and reshaping a galaxy over time, the significance of the third mechanism continues to be debated.   In particular, it has not been empirically demonstrated how effective gravitational torques (induced by non-axisymmetric features like bars and spiral arms) are at redistributing angular momentum, which can lead to re-shaping the dynamics and stellar density profile of a galaxy over time.  

In this present study we develop a framework on how to combine simulations, 'stellar mass density' images of spiral galaxies and sample averaging to address this question.  Specifically, we examine 24 galaxies with photometry from SDSS in order to assess how strong gravitational torques are at any given instant on the stellar distribution.  Unless otherwise stated, we consider torques due to the stellar distribution acting solely on the stellar distribution.   This modest galaxy sample serves to a) determine the limitations of both the method and the data; b) determine if gravitational torques are significant for a set of individual galaxies and c) explore how samples of galaxies should be 'stacked' in order to establish an average measure over time of the secular evolution for galaxy disks.  Subsequent papers will include an analysis of a much larger sample of galaxies.  In \S 2 we describe our motivation and in \S 3 we describe how the torques are calculated.  Tests completed with N-body/SPH simulations are outlined in \S 4.  The observations and analysis are described in \S 5 and \S 6 and \S 7 summarizes the conclusions. 

\section{Background \& Motivation}

Stellar bars and spiral arms are common in and eponymous for disk galaxies. Two key questions present themselves for the study of bars and spiral arms: a) why do these features exist and look the way they do? and b) given that we observe these features, what instantaneous effect do they have?  In the present study, we focus on the second one.  Previous studies have attempted to calculate the strength of bars and spiral arms and their role in angular momentum transfer.  The early analytical calculations of Lynden-Bell \& Kalnajs (1972) showed that a trailing spiral structure will lead to an overall outflow of angular momentum, and observations indicate that most spirals are trailing ({\it e.g.} de Vaucouleurs, 1958, Pasha \& Smirnov, 1982 and Pasha 1985).   Bertin (1983) derived a time scale for galaxy evolution due to a spiral mode from the angular momentum contained in the disk and the angular momentum flux.  His work suggested a time scale which was an order of magnitude greater than a Hubble Time.   However, he did not use spiral arm parameters from observed galaxies, but rather chose what he presumed to be reasonable values for an average spiral.  

Gnedin et al. (1995) (hereafter G95) pursued this further, using direct observations of M100 to construct a deprojected stellar mass map for this galaxy and calculate the torques.  They found that if the instantaneously observed torques in M100 were to act over long periods, they would redistribute angular momentum on a timescale of 5-10 Gyr, i.e. they would re-shape the stellar distribution within a Hubble Time.

Torques exerted by the stars also act on the (cold) gas.  Several studies have explored the effects of such gravitational torques exerted by stars on gas and used them to measure gas inflow rates in galaxies ({\it e.g.} Jogee et al. 2005, Hunt et al. 2008, Haan et al. 2009).  In particular, Haan et al. (2009) has shown that gravitational torques are extremely efficient at redistributing cold gas and can act on a timescale of a few dynamical periods, $5 \times 10^{8}$ Gyr.  

The problem in studying this effect in individual galaxies lies in the fact that we can only measure the instantaneous torques, but are interested in their long-term ($t_{\rmn{dyn}} \ll t < t_{\rmn{Hubble}}$) secular effect.  Therefore, one would be forced to assume that the spiral and bar patterns seen today are typical of the past and future.  However, for any given galaxy this is likely not the case.  

Many people have turned to simulations since they allow us to follow a galaxy over time.  Indeed many simulations (Zhang, 1996, 1998, 1999, Sellwood \& Debattista, 2000 and Athanassoula, 2002, 2003) have tried to address the strength and effect of bars and spiral arms over time.  In a series of theoretical papers, Zhang showed that gravitational torques lead to stars being funneled inward within the corotation radius, moving outward beyond; for large spiral galaxies, $R_{co-rot}$ $\approx$ $3~r_{\rmn{exp}}$  (Kranz et al., 2003).  The mass redistribution was shown to imply some development of the Hubble Sequence (Zhang, 1996, 1998, 1999).  

Following the analytical work of Lynden-Bell and Kalnajs (1972), Athanassoula (2002) used simulations to study the angular momentum transfer at resonances, specifically focusing on how the halo affects the amount of angular momentum transferred and subsequently the growth of the bar.  These studies confirmed that within the inner Lindblad resonance angular momentum is lost, while at corotation and the outer Lindblad resonance angular momentum is gained. 
 
The simulations of Sellwood \& Binney (2002) and Ro{\v s}kar et al. (2008) showed that radial mixing due to angular momentum transfer occurs even in the case where little change is seen in the mass density profile.  Disk stars were shown to migrate considerably over time due to resonant scattering.  Therefore, the translation of angular momentum transport to changes in the stellar surface mass density profile is complex.

In general, angular momentum redistribution has been shown to lead to higher central mass density concentration, to the development of a two-component profile and the evolution of the inner disk scale length over time toward higher values (Debattista et al. 2006, Foyle et al. 2008).  

While simulations and analytical studies have certainly provided much insight into the role of resonances and angular momentum transfer, for our study we wish to focus on the strength of torques on stars by the stellar distribution itself and how these torques contribute to angular momentum transfer.  For such a study, numerical  simulations of disk galaxies are not ideal.  The mass distribution and in particular the non-axisymmetric components  (i.e. bars and spiral arms) are the key to the strength of the torques.  Early numerical simulations have shown that spiral arms are transient features that dissipate after $\approx$ 10 rotations (Sellwood \& Carlberg, 1984).  Even modern simulations that include many more physical processes confirm the short-lived and recurrent nature of spiral arms (Governato et al. 2007).    It is unclear, however, whether spiral arms are short-lived in nature.  Lin \& Shu (1964) and Bertin et al. (1989) have proposed long-lived spiral density waves.  

Furthermore, an even greater challenge is that disk galaxy simulations embedded in a cosmological context are far from being able to produce realistic galaxies that match both the size-velocity-luminosity and color-luminosity relations (Courteau et al. 2007;  Dutton et al 2007).   These galaxies typically produce very high central mass concentrations due to angular momentum transfer ({\it i.e.} Navarro \& Benz 1991; Navarro \& White 1994; Navarro \& Steinmetz 1997).  In recent years, these cosmological simulations have greatly improved by including more physical processes and additional resolution ({\it i.e.} Thacker \& Couchman 2001; Somerville 2002; Governato et al. 2007), but the problem is not fully resolved.   In light of this, one can not rely on simulations alone to estimate the actual secular evolution rate in observed galaxies.  Indeed, depending on the duration for which a simulation has run,  simulations may show that the torques are either much stronger or weaker than those seen in nature.  In our present study, we focus on a purely empirical assessment of the exerted torques and do not address the detailed physics of angular momentum transport.  While we employ simulations, these are only used to provide mock observations to test various steps in our method and examine the effects of 'non-observables'.

In order to assess how important gravitational torques actually are for shaping galaxies over long times, studying individual galaxies is not ideal.  In this case, in fact, we are forced to assume that for both the angular momentum, $J$ and the torque, $\Gamma$, we have $J(R,t_{\rmn{now}}) \approx J(R,t_{\rmn{later}})$ and $\Gamma(R,t_{\rmn{now}}) \approx \Gamma(R,t_{\rmn{later}})$ respectively.   G95 made this assumption, but acknowledged that the structure of the stellar distribution will change in any galaxy's past or future.   For an ensemble of galaxies whose overall population properties are evolving slowly, we can say that $\langle J(R,t_{\rmn{now}})\rangle_{\rmn{sample}} \approx \langle J(R,t_{\rmn{later}})\rangle_{\rmn{sample}}$ and $\langle\Gamma(R,t_{\rmn{now}})\rangle_{\rmn{sample}} \approx \langle \Gamma(R,t_{\rmn{later}})\rangle_{\rmn{sample}}$. Thus, we propose to bypass this problem by averaging over a large sample of similar-sized galaxies not chosen based on similar spiral patterns or bar strengths.   Bar and spiral arm features are not recent phenomena and many studies have found that the prevalence of bars and spiral features has not changed very much since z $\sim$ 0.7, a period that corresponds to $\sim 20$ dynamical times for the half-mass radius of large disks (Jogee et al. 2005, Barden et al. 2005, Marinova \& Jogee 2007 and Barazza et al. 2008).  While some studies, including Sheth et al. (2008), found a decrease in bar fraction with redshift particularly for low-mass blue spirals, this effect may be due to a decrease in resolution and obscuration by star formation and dust (Marinova \& Jogee 2007). Thus, even if the bar fraction is not constant, it may only exhibit a moderate decline of a factor of 2.  Our stacked ensemble is an average for low redshift galaxies and should not be considered an average extending to high ($z > 1$) redshift.

\section{Determining the Angular Momentum Flow Timescale from Photometric Data}

The first step of our experiment is to determine the instantaneous rate of angular momentum flow for any given spiral galaxy disk at each radial position. Ideally we would have the full 3D potential due stars, gas and dark matter, $\tilde{\Phi}_{\rmn{tot}}(r,\varphi,\theta)$ and the density distribution of the stars $\rho(r,\varphi,\theta)$.  However, in practice we can at best have information on the projected stellar mass density and we employ an effectively softened potential that can be based on an observational estimate of the stellar mass surface density, $\Sigma_{\rmn{\star}}(r,\varphi)$.  In \S 4 we will use simulations to test how well this works, by looking at the effects of dark matter and gas and the difference between employing the 3D and 2D potential.

There are two important steps in calculating  angular momentum transfer from
photometric data.  The first step involves determining a map of the stellar
mass distribution from surface photometry. We have looked at this problem
closely (Zibetti et al. 2009; hereafter Z09), and we will summarize how this is done in \S
5.2; for now we will assume that we have the stellar mass distribution known.  

The second step requires calculating the torques on the outer mass
distribution ($r > R$) due to the inner mass distribution ($r < R$) in order
to determine an angular momentum flow rate across radius R (see G95).  We determine the $z$-component of the torque at each position $R$ by summing the gravitational forces on the outer region due to the inner region and taking the cross product with the position vector: 
\begin{eqnarray}
\Gamma(R)=
\end{eqnarray}
\begin{eqnarray*}
 {}{}{}{}{}G \int_{r_{\rmn{1}} > R} d^{2}{\bf r_{\rmn{1}}}  \Sigma(r_{\rmn{1}}) \int_{r_{\rmn{2}} < R} d^{2}{\bf r_{\rmn{2}}} \times \Sigma(r_{\rmn{2}}) \frac{({\bf r_{\rmn{1}}} \times {\bf r_{\rmn{2}}})_{\rmn{z}}}{[ |{\bf r_{\rmn{1}}}-{\bf r_{\rmn{2}}}|^{2} +\epsilon^{2}]^{\frac{3}{2}}},
\end{eqnarray*}
where ${\bf r}$ is the projection vector on the equatorial plane and
$\Sigma(r)$ is the stellar mass per unit area.  We use a factor $\epsilon$ to
account for disk thickness.  G95 examined in detail how varying the treatment
of disk thickness will affect the torque calculation and found it to be of
little import. Thus, we adopt $\epsilon = \frac{0.7~r_{\rmn{exp}}}{12}$ and examine the accuracy of this choice by testing it with the simulations (see \S 4).

The rate of angular momentum flow can be derived by comparing the torque across radius $R$, $\Gamma(R)$, to
the total angular momentum interior to $R$, $J(R)$, which is  expressed as:
\begin{eqnarray}
J(R)=\int_{0}^{2 \pi} d\theta \int_{0}^{R} rdr rV_{\rmn{c}} \Sigma(r,\theta)  ,
\end{eqnarray}
where $V_{\rmn{c}}$ is the rotational velocity.  Using the Tully-Fisher relation and an $arctan$ rotation curve
parameterization (Courteau, 1997)  we are able to construct approximate rotation curves
for each galaxy from photometry alone.  The sign of rotation is determined by assuming the galaxies
have trailing spiral arms ({\it e.g.} de Vaucouleurs, 1958, Pasha \& Smirnov, 1982 and Pasha 1985).  We define a torque-driven rate of angular
momentum flow,
\begin{eqnarray}
\nu_{\rmn{inflow}}(R)= \frac{\Gamma(R,t)}{J(R,t)},
\end{eqnarray}
which has dimensions of time$^{-1}$.  We can assess if torques are strong enough to be significant if the product of $\nu_{\rmn{inflow}}$ and a Hubble Time is greater than unity.  We can re-interpret $\nu_{\rmn{inflow}}$ qualitatively as the rate of mass flow in the opposite
direction.   We determine $\nu_{\rmn{inflow}}$ at each position in the disk and then scale by disk scalelength.  The scaled values of $\nu_{\rmn{inflow}}$ are then used to create a sample average profile by stacking the plots.  This allows us to determine  the mean rate of matter inflow for present day spiral galaxies.

However, the total angular momentum flux need not depend on the gravitational torques alone.  In the case of a long lived spiral structure, density wave theory predicts that angular momentum will also be advected (also known as 'lorry' transport).    Advective transport is due to the bulk motion of the fluid and does not lead to matter flow.  Depending on the pitch angle of the spiral and its longevity, advective transport may overwhelm the gravity torque by transporting angular momentum inward (G95 and Binney \& Tremaine, 2008).    However, almost all simulations show that spiral features are not long-lived but are recurrent transient features (Sellwood 2010, Sellwood \& Binney 2002, Governato 2007, etc.).  There is also a substantial amount of observational evidence that support transient spirals (Sellwood 2000, Merrifield et al 2006, etc.).   If this is the case, the role of advective transport would be small as its effects are greatest in the case of a long-lived spiral structure with a constant pattern speed.  There is no way to directly observe the strength of advective transport and so, our estimation here cannot account for it.

Also our measure of angular momentum flow need not translate straightforwardly into matter flow.  Our torque measure represents the net change of the angular momentum distribution at each radial position.  Changes in the angular momentum distribution may cause changes in the mass density profile of the disk, provided the stellar orbits are not too eccentric.  Loss of angular momentum inside corotation causes the home radii of the stars to get smaller.  If the orbits of the stars are very eccentric, the stars spend a disproportional amount of time away from the home radius and thus the mass profile changes are weakened (Sellwood \& Binney, 2002).  The extent to which our measure of angular momentum flow can be translated into changes in the mass density profile is beyond the scope of this paper, but is certainly essential to understanding secular evolution.  For our discussion, we caution the reader that we do not treat these complexities.

\section{Testing Torque Estimates using Disk Galaxy Simulations}

There are several challenges involved in using observations to derive an
accurate torque map of a galaxy, where simulations can be used to test and
refine the method.  For instance, the gravitational potential of an observed
galaxy cannot be strictly derived from observables (especially without
detailed kinematics) and so one must make assumptions about the matter distribution.  Here, we first use simulations to test how the accuracy of torque maps from
projected stellar distributions is affected by the vertical extent of the stellar distributions.  Second, we explore the role of gas and dark matter in exerting torques on
stars. Making mass maps of the gas and dark matter component of the galaxies
is difficult or impossible, respectively.  Simulations with live halos and gas
will allow us to determine the significance of their torques on the stars and potentially derive a correction factor for their effects.

In the following sections we show that a map of the projected stellar mass enables a sensible way to estimate the total torque on the stars; if one wants a correction for dark matter torques, we offer one.

\subsection{Simulations}

\begin{figure*}

\centering

 \includegraphics[width=80mm]{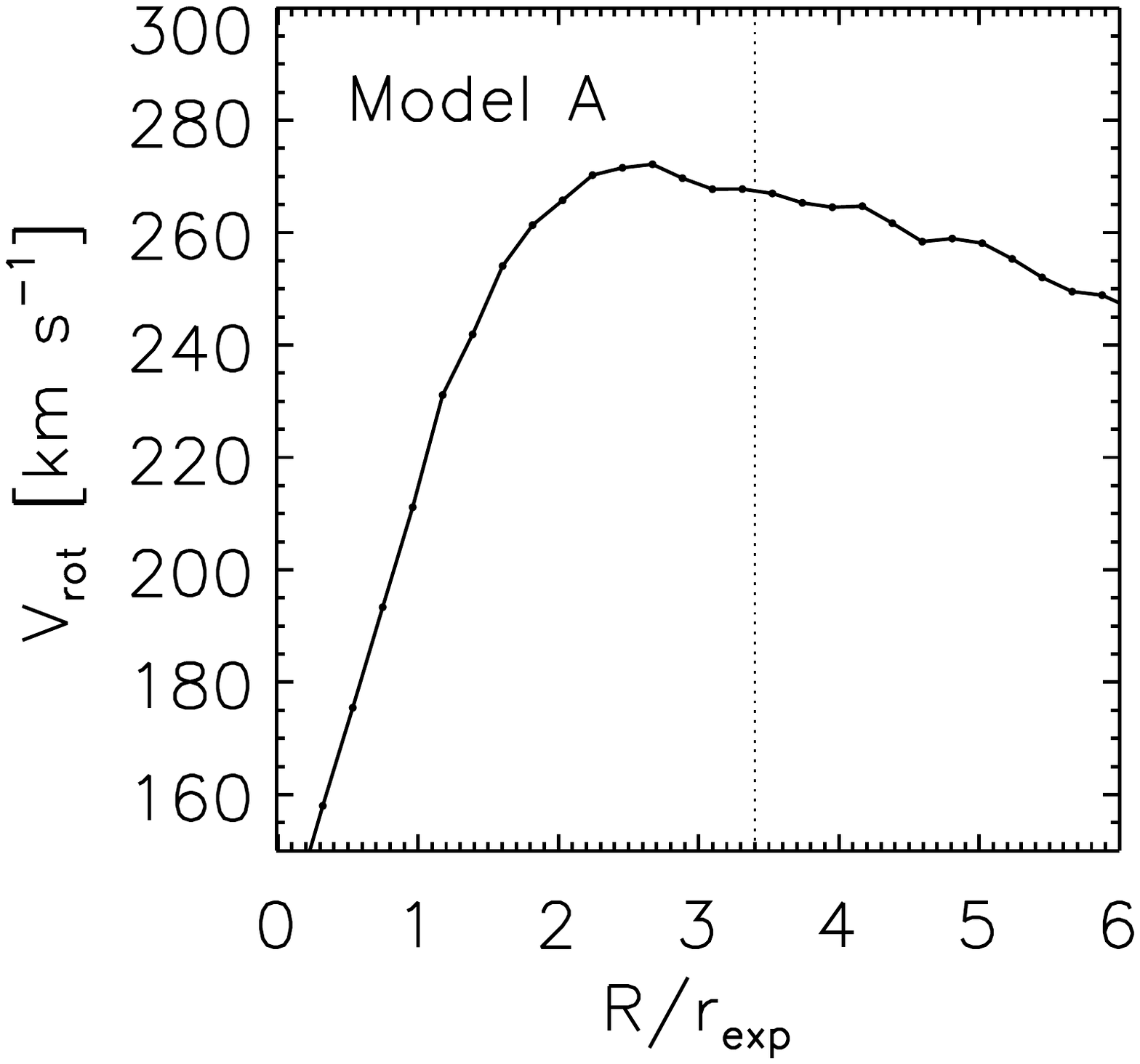}
 \includegraphics[width=80mm]{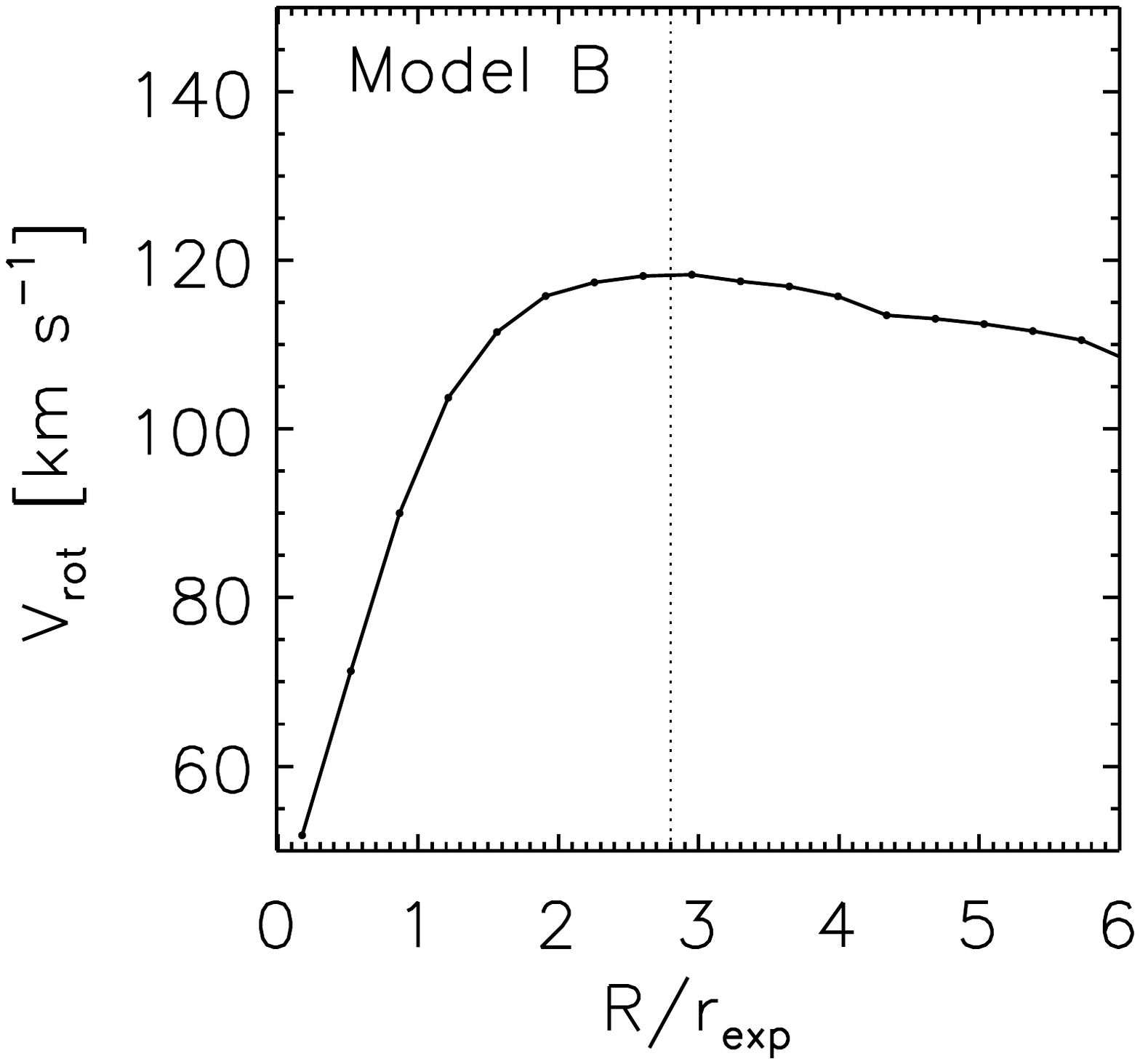}
\rm

\caption{Rotation curve for models A and B after 2 Gyr of evolution.  The dashed line shows the position of corotation of the spiral pattern in the model. }

\label{model_rot}

\end{figure*}

\begin{figure*}

\centering

 \includegraphics[width=140mm]{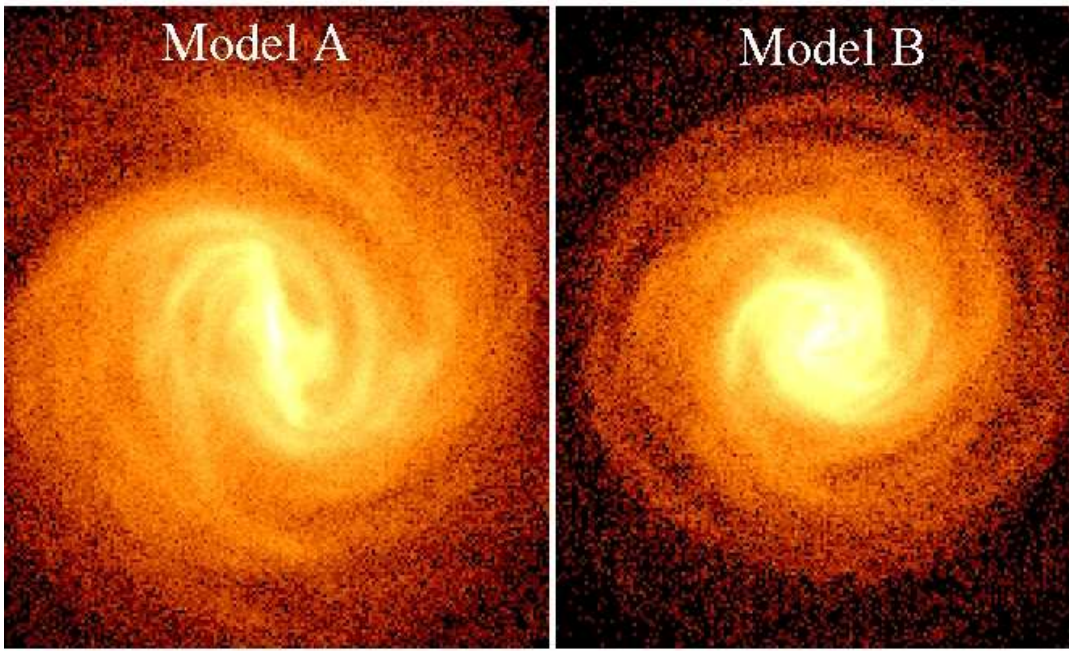}

\rm

\caption{Snapshots of the stellar component of models A (left) and B (right),
  illustrating the stellar surface mass density 2 Gyr after the symmetric
  simulation start, when they showed a strong bar/spiral arm
  structure.  Rotation is clockwise in both models, with spiral features trailing.  }

\label{images}

\end{figure*}

We use two simulated isolated model galaxies each with over 1.4 million particles ($N_{\rmn{halo}}=6 \times 10^{5}$, $N_{\rmn{disk}}=4 \times 10^{5}$, $N_{\rmn{gas}}=4 \times 10^{5}$) from the work of Foyle et al (2008) in order to conduct our tests.  The simulations rely on the GADGET-2 code (Springel 2005) with star formation and feedback as described in Springel \& Hernquist (2003).  Table 1 lists the properties of the selected galaxies including the halo spin parameter, $\lambda$, halo concentration, $c$, mass or virial velocity, $V_{\rmn{200}}$, disk mass fraction, $m_{d}$, disk scale length, $r_{\rmn{exp}}$, the peak velocity of the rotation curve $V_{\rmn{peak}}$  and the position of the velocity peak in terms of disk scale length ($R_{\rmn{peak}}/r_{\rmn{exp}}$).  Fig.~\ref{model_rot} shows the rotation curve for the two model galaxies.  The dashed line marks the position of corotation of the spiral pattern.  Corotation was calculated by measuring the angular speed of the spiral pattern and comparing with the rotation speed at that position.  Fig.~\ref{images} shows images of the stellar mass density of the models.  The models began with straight exponential profiles and were run for 10 Gyr.  We chose snapshots after roughly 2 Gyr where the bars and spiral arms had formed, as these features are most relevant for the gravitational torques.  We do not attempt to simulate realistic disk galaxies here.  We require only 'plausible' models in order to test the method used on our observed galaxy sample.  Further details of the simulations and star formation model are described in Foyle et al. (2008).

In the simulations, we calculated the torques due to the full potential
throughout the galaxy and consider the potential due to the stars, gas and
halo. This is somewhat different from the observationally robust approach
described in \S 3, where we consider the torques on the outer parts due to the
stellar distribution in the inner parts.   We also calculate the total angular momentum in an annulus as opposed to the total angular momentum within $R$.  In order to distinguish these
calculations in the simulations from those described in \S 3, we label them as $\Gamma_{\rmn{z}}$ and $J_{\rmn{z}}$, as opposed to $\Gamma$ and $J$ (see Eq. 4).  In order to estimate the strength of the simulation torques we divide $\Gamma_{\rmn{z}}(R)$ at each radial position by the total angular momentum in an annuli centered at $R$.  We multiply by $\Delta t$, which is set to a time span of 1 Gyr.  1 Gyr is longer than a dynamical time, but much less than a Hubble Time, allowing us to ascertain the present strength of the torques in the simulation.

\begin{table}

\caption{Model Properties}

\begin{tabular}{cccccc}

\hline\hline
& Model A & Model B \\
$\lambda$ & 0.08 & 0.08 \\
$c$ & 15 & 10 \\
$V_{\rmn{200}}$ [km s$^{-1}$]  & 160 & 80 \\
$m_{d}$ & 0.1 & 0.1 \\
$r_{\rmn{exp}}$ [kpc] & 4.68 & 2.88 \\
$V_{\rmn{peak}}$ [km s$^{-1}$] & 272 & 118\\
$R_{\rmn{peak}}/r_{\rmn{exp}}$ & 2.67 & 2.95\\
\hline


\end{tabular}

\end{table}



\begin{figure*}

\centering

 \includegraphics[width=140mm]{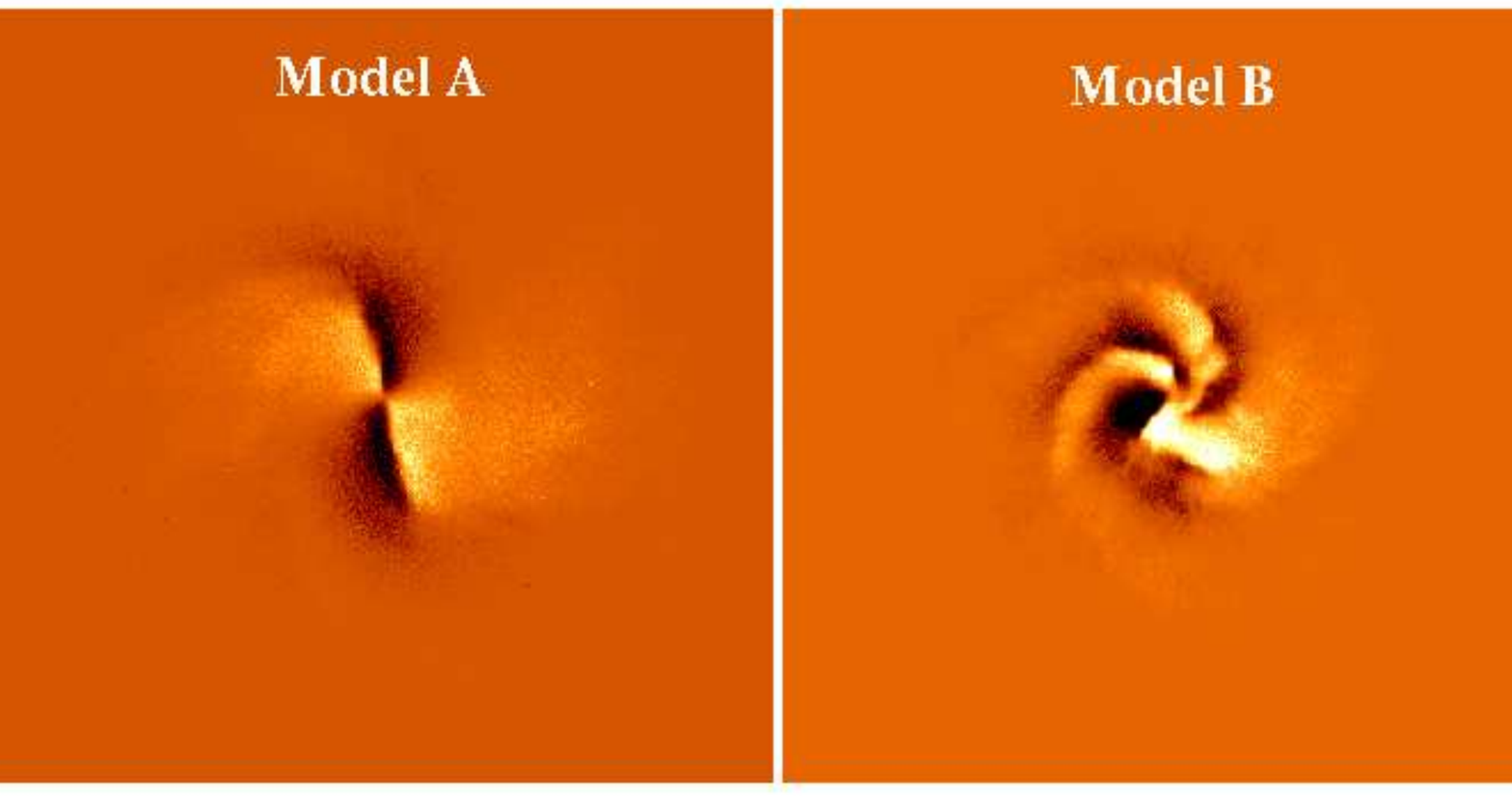}

\rm

\caption{Snapshots torques on stars by stars of models A (left) and B (right) after 2 Gyr of evolution when they showed a strong bar/spiral arm structure.  Model A shows considerable symmetry in comparison to model B.}

\label{imagestorques}

\end{figure*}



\subsection{Projected Mass Densities and the 2D Potential}

Previous works (Buta et al. 2004) have attempted to derive the 3D potential of the stellar disk
directly by using Fourier techniques and assumptions about the vertical height
distribution.  Following G95, we choose to use a softened 2D potential and
then calculated the torques directly by determining the pixel scale and adding a correction term based on the disk scale length of the galaxy ($\epsilon=\frac{0.7 r_{\rmn{exp}}}{12}$).  G95 used a similar softening approach, but derived the potential using Fourier techniques.

\begin{figure*}

\centering

 \includegraphics[width=80mm]{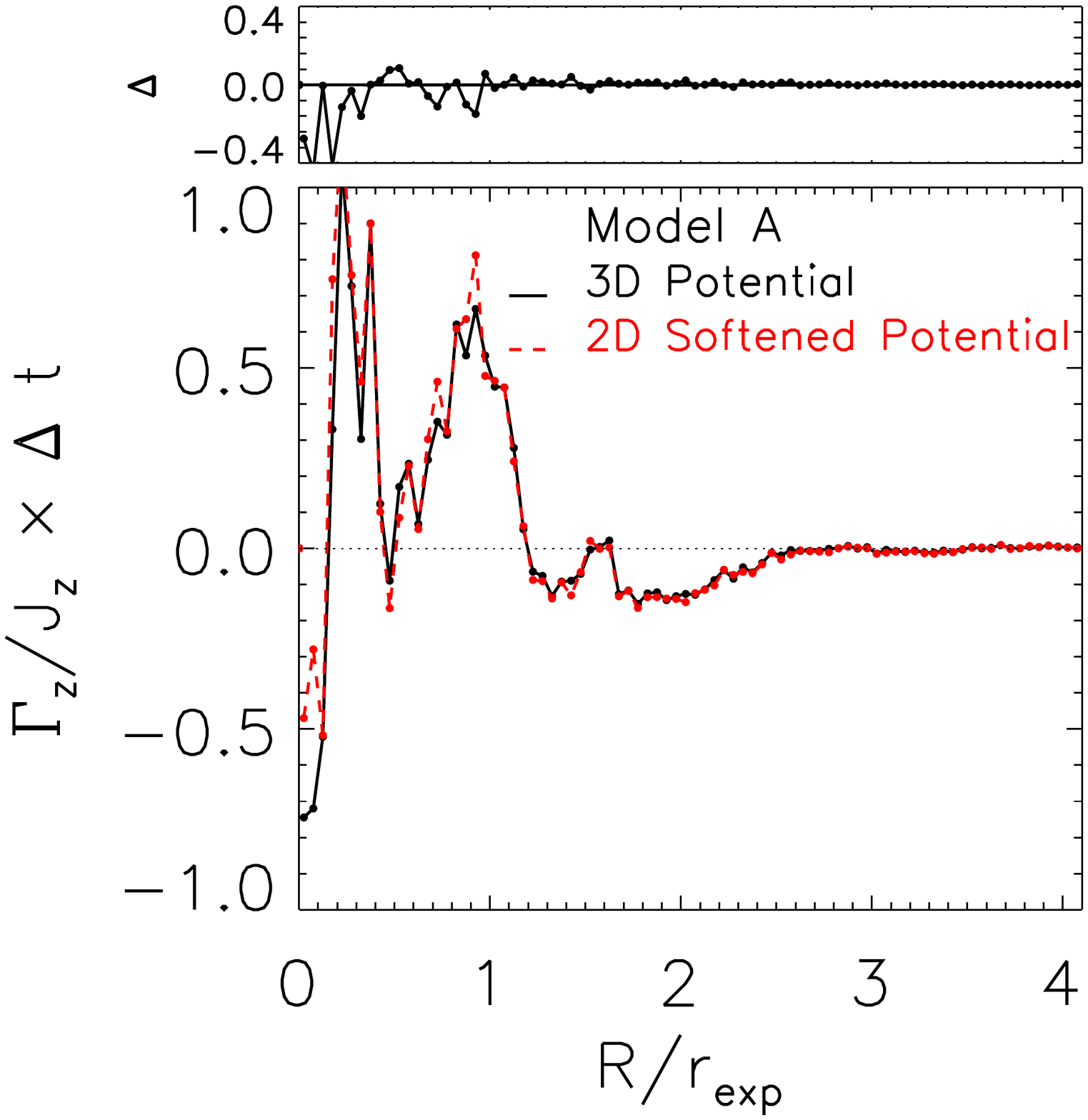}
 \includegraphics[width=80mm]{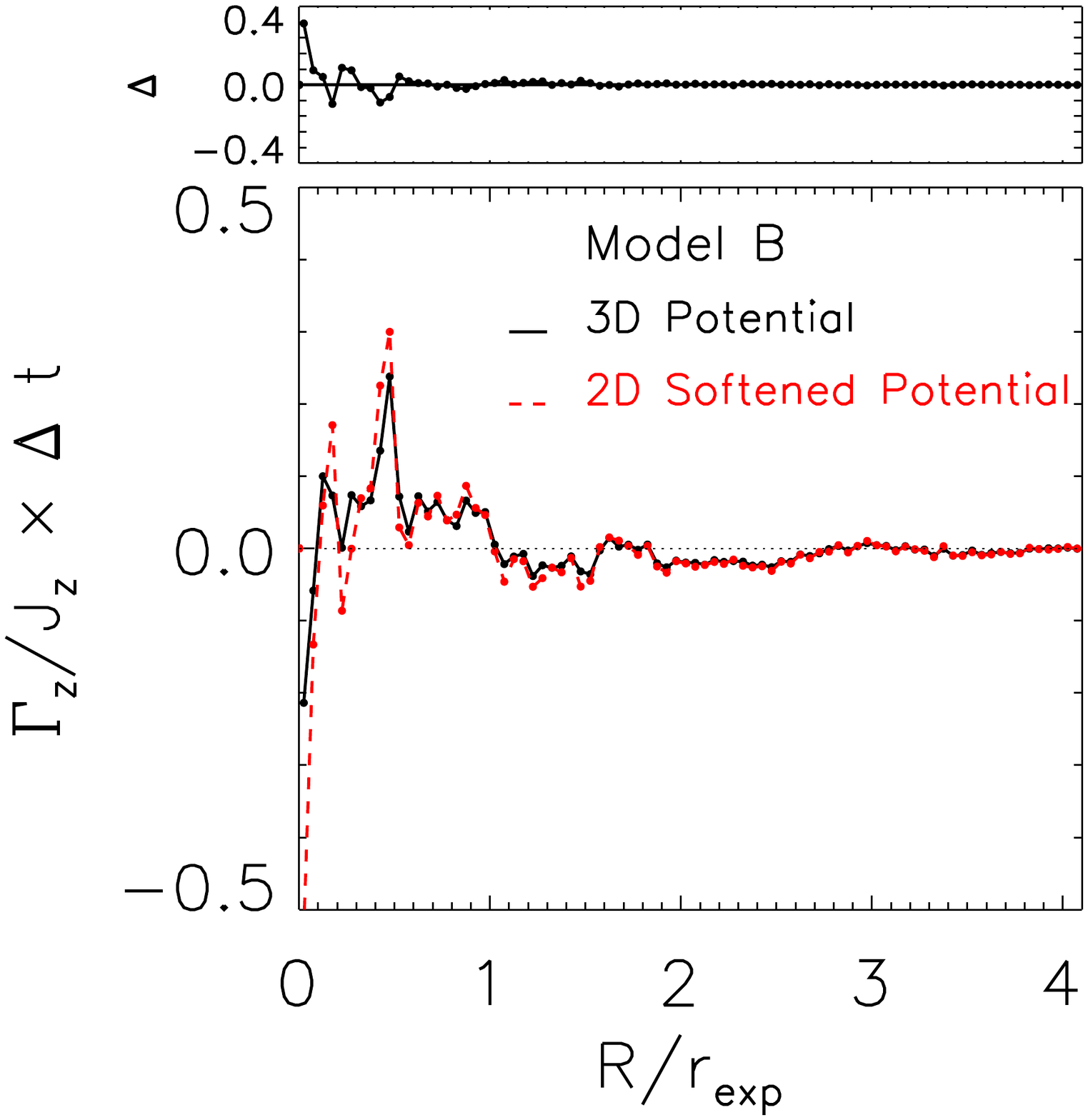}

\rm

\caption{Accuracy of torques from projected mass densities in comparison to
  the 3D potential: the azimuthally
  averaged scaled torques due to the stellar particles where the distances are
  calculated from all three components (black line) and where the distances
  are calculated using only the $x$ and $y$ components and a correction factor
  for the $z$-component (red line) after 2 Gyr of evolution.  The torques, $\Gamma_{\rmn{z}}(R)$, are divided by the total angular momentum in each annulus and multiplied by $\Delta t$, which is set to 1 Gyr, giving a dimensionless estimate of the strength of the torques.  Using a softened
  2D potential to calculate the torques yields an accurate estimate of those
  calculated using the full 3D potential.  The upper panel shows the
  difference between the 2D softened potential and the 3D potential versus
  radius.  Only a few peaks are stronger with the 2D potential and the average deviation between the two is 4\%. }

\label{proj}

\end{figure*}

Specifically, our torques are calculated as:
\begin{eqnarray}
\Gamma_{\rmn{z,pixel}}=x_{\rmn{pixel}}F_{\rmn{y,pixel}}-y_{\rmn{pixel}}F_{\rmn{x,pixel}},
\end{eqnarray} 
where
\begin{eqnarray*}
F_{\rmn{x,pixel}}=m_{\rmn{pixel}} \sum_{j=0}^{n_{\rmn{p}}} \frac{G m_{\rmn{j}} (x_{\rmn{pixel}}-x_{\rmn{j}}) }{((x_{\rmn{pixel}}-x_{\rmn{j}})^{2} + (y_{\rmn{pixel}}-y_{\rmn{j}})^{2} +\epsilon^{2})^{3/2}}
\end{eqnarray*} 
and
\begin{eqnarray*}
F_{\rmn{y,pixel}}=m_{\rmn{pixel}} \sum_{j=0}^{n_{\rmn{p}}} \frac{G m_{\rmn{j}} (y_{\rmn{pixel}}-y_{\rmn{j}}) }{((x_{\rmn{pixel}}-x_{\rmn{j}})^{2} + (y_{\rmn{pixel}}-y_{\rmn{j}})^{2} +\epsilon^{2})^{3/2}},
\end{eqnarray*} 
where the sum index $j$ runs through all pixels in the 2D image.  Fig.~\ref{imagestorques} shows the $\Gamma_{\rmn{z,pixel}}$ for the torques due to the stellar distribution on the stars.  Model A shows considerable symmetry in comparison to model B.  When calculating the torques on stars by stars, the result is strongly affected by the symmetry of the distribution.  When one azimuthally averages, considerable cancellation will occur in symmetric distributions.
In Fig. ~\ref{proj}, we plot the azimuthally averaged value of $\Gamma_{\rmn{z}}(R)/J_{\rmn{z}}(R) \times \Delta t$ for annuli of 0.05$r_{\rmn{exp}}$.  $J_{\rmn{z}}$ is the total angular momentum within the annuli.  We see that, in the inner parts, the torques are strong and act on timescales significantly less than a Hubble Time (recall $\Delta t = 1$ Gyr).   We see that for both model A and B, the
softened 2D force (red dashed line) is indeed an accurate estimate of the full in-plane
component of the 3D force (black solid line),  with the  average deviation between the two of
only 4\%.  Therefore, knowing only the projected stellar mass distribution is not a serious limitation.



\subsection{Torques on Stars from Stars versus Torques from the Gas \& Dark Matter}

Observations of stellar mass maps, as can be made available for large samples,
will only allow us to measure the torques due to the stellar mass
distribution.  We use the simulations to test how significant we should expect
the torques to be due to the gas and dark matter.  In Fig. ~\ref{comp}, the torques, $\Gamma_{\rmn{z}}(R)$, are divided by the total angular momentum in each annulus, $J_{\rmn{z}}(R)$, and multiplied by $\Delta t$, which is set to 1 Gyr, giving a dimensionless estimate of the strength of the torques.  In Fig.~\ref{comp} we consider the torques 
 on the
stars due to the stars (blue), gas (red) and dark matter (cyan).  We see that
the torques on the stars due to the gas are quite small and can be neglected in comparison to the other two.   The dark matter torques are significant, but almost always act in the same direction as the stellar torques.  Thus, a measure of the stellar torques gives us at least a minimum measure of the total torque.

Fig.~\ref{cumul} shows the cumulative distribution of the ratio of the torques due to the stars versus the total torques of the stars and dark matter.  The particles in the inner disk scale lengths have torques which are predominately due to the stellar distribution.  Thus, if the torques are significant, we can conclude that the stellar torques dominate the total torques and, if anything, the dark matter torques enhance the stellar torques.  

We can also use the simulations to derive a correction factor for the dark
matter torques.  We determine a linear enhancement function, by
connecting the two median points in the inner ($r<2r_{\rmn{exp}}$) region and outer ($2r_{\rmn{exp}}<2<5r_{\rmn{exp}}$)
region of the disk.  Fig.~\ref{enhance} shows the fit for model A, which was selected as it had the
strongest torques of the two models.

If we calculate the torque as we will for the observed galaxies, that is the torque on the outer stellar distribution ($r > R$) due to the inner stellar distribution ($r < R$) as described in \S 3, we see that that for both models the peak of the torque curve lies interior to $1.5r_{\rmn{exp}}$, implying that the highest rates of inflow will be in the inner regions of the disk (see upper panel of Fig.~\ref{tgnedin}). For both galaxies the inner regions show predominately angular momentum outflow.    We also note that the peak value for the torque of model A is much greater than model B.  This is mostly due to the fact that model A is a more massive galaxy with a $V_{\rmn{200}}$ of 160 km s$^{-1}$ as opposed to 80 km s$^{-1}$ for model B and that the torque scales by the square of the mass. The quantity we need to consider when evaluating the strengths of the gravitational torques is the timescale for angular momentum flow.  The timescale is given by $\nu_{\rmn{inflow}}$ which is the ratio of the torque, $\Gamma(R)$, to the angular momentum, $J(R)$, as calculated in Eq. 2.  While the amplitude of the torque curves varies considerably between the models, a comparison of the timescales shows that these are roughly similar (see lower panel of Fig.~\ref{tgnedin}).  

The simulations have allowed us to show that the projected stellar mass can be used to accurately calculate the torques and the torques due to the stellar distribution on the stars represent at least a minimum measure of the total torques due to the stellar, gas and dark matter distributions.  We have been able to derive an enhancement function, which can be applied to the observed stellar torques to account for the torques due to the dark matter.  Again, we wish to stress that when we compare the torques on the stars due to the stellar, gaseous and dark matter distributions, we have been calculating the torques due to the full potential of these distributions.  This differs to what we do observationally (see \S 3).  However, if the torques due to the dark matter are less than that of the stars when considering the full potential, it follows that they will also be less when considering the torque flow through a given radius.  


\begin{figure*}

\centering

 \includegraphics[width=80mm]{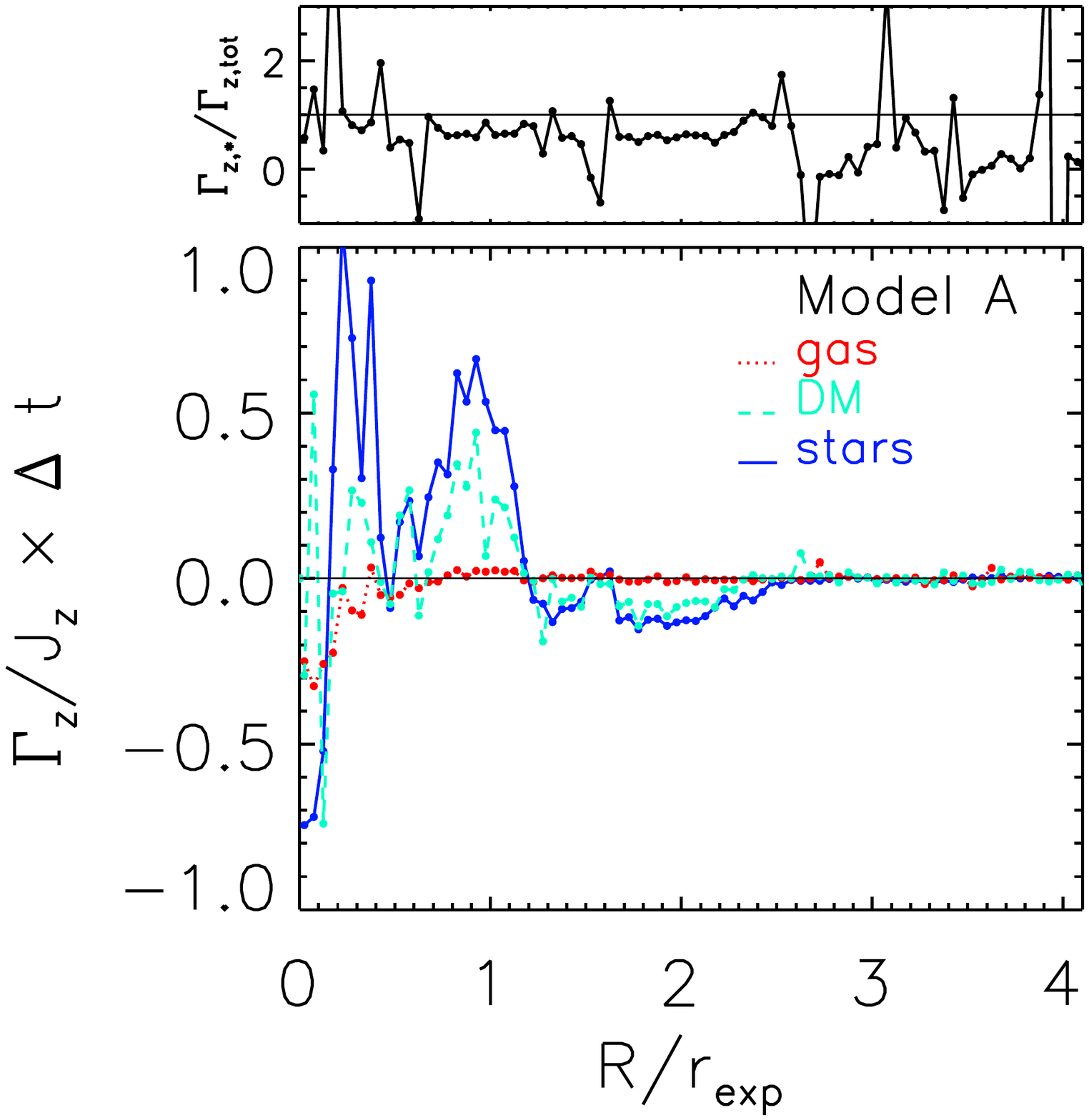}
  \includegraphics[width=80mm]{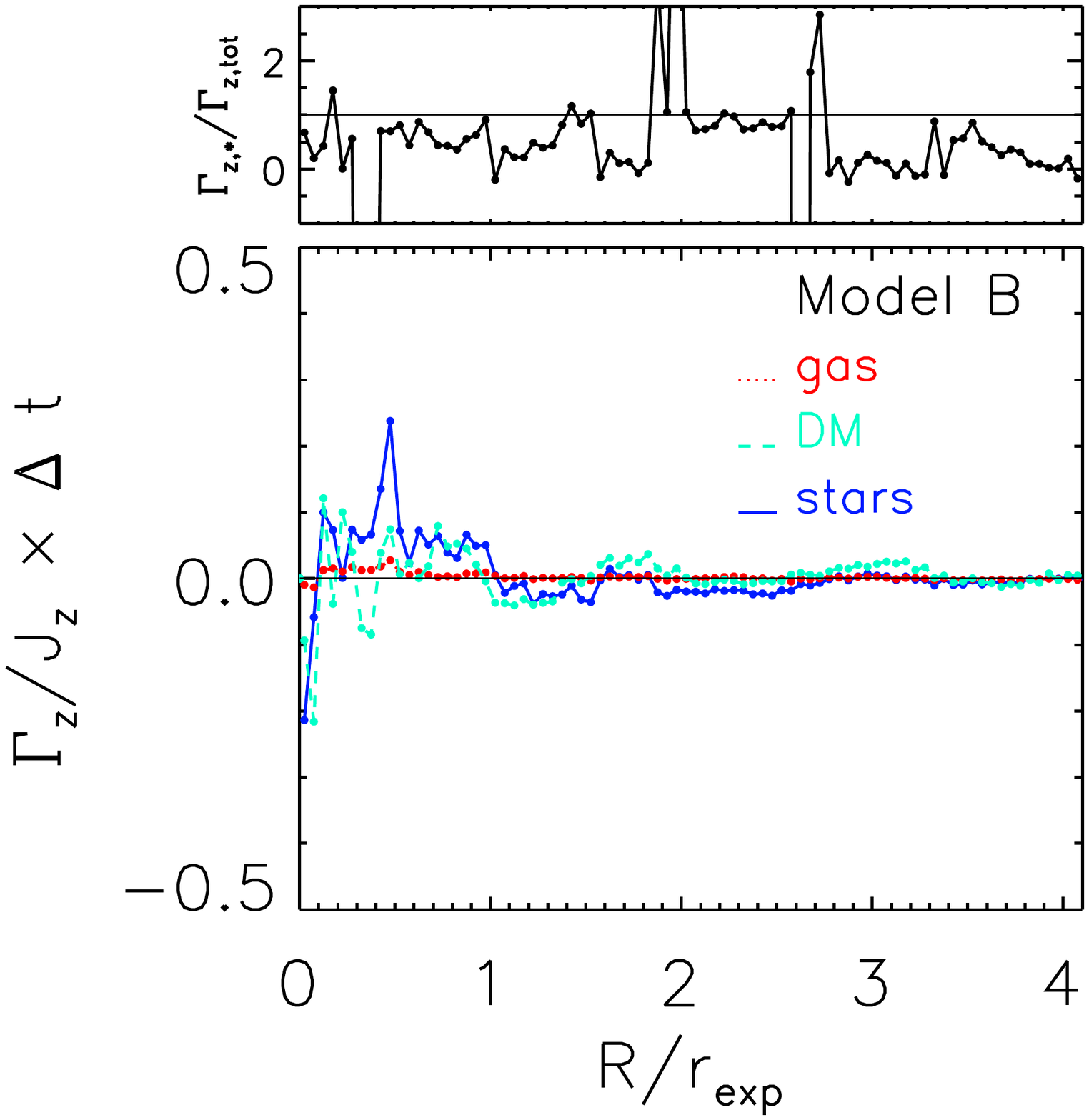}

\rm

\caption{The relative importance of torques exerted by stars, gas and dark
  matter: a comparison of the scaled azimuthally averaged torque on the star
  particles due to the star (blue), gas (red) and dark matter (cyan) particles
  for models A and B. Model A shows considerably stronger torques than model
  B.  In both cases the dark matter torques are weaker than that of the
  torques due to stellar distribution.  Furthermore, the torques are such that
  they almost always act in the same direction.  The torques due to the dark
  matter distribution serve to enhance the torques due to the stellar
  distribution.  The upper panel shows the ratio of the stellar torques to the
  total torques.  In only a few cases do the directions differ and over much
  of the disk the stellar torques make up the majority of the total
  torques.}

\label{comp}

\end{figure*}



\begin{figure*}

\centering

  \includegraphics[width=80mm]{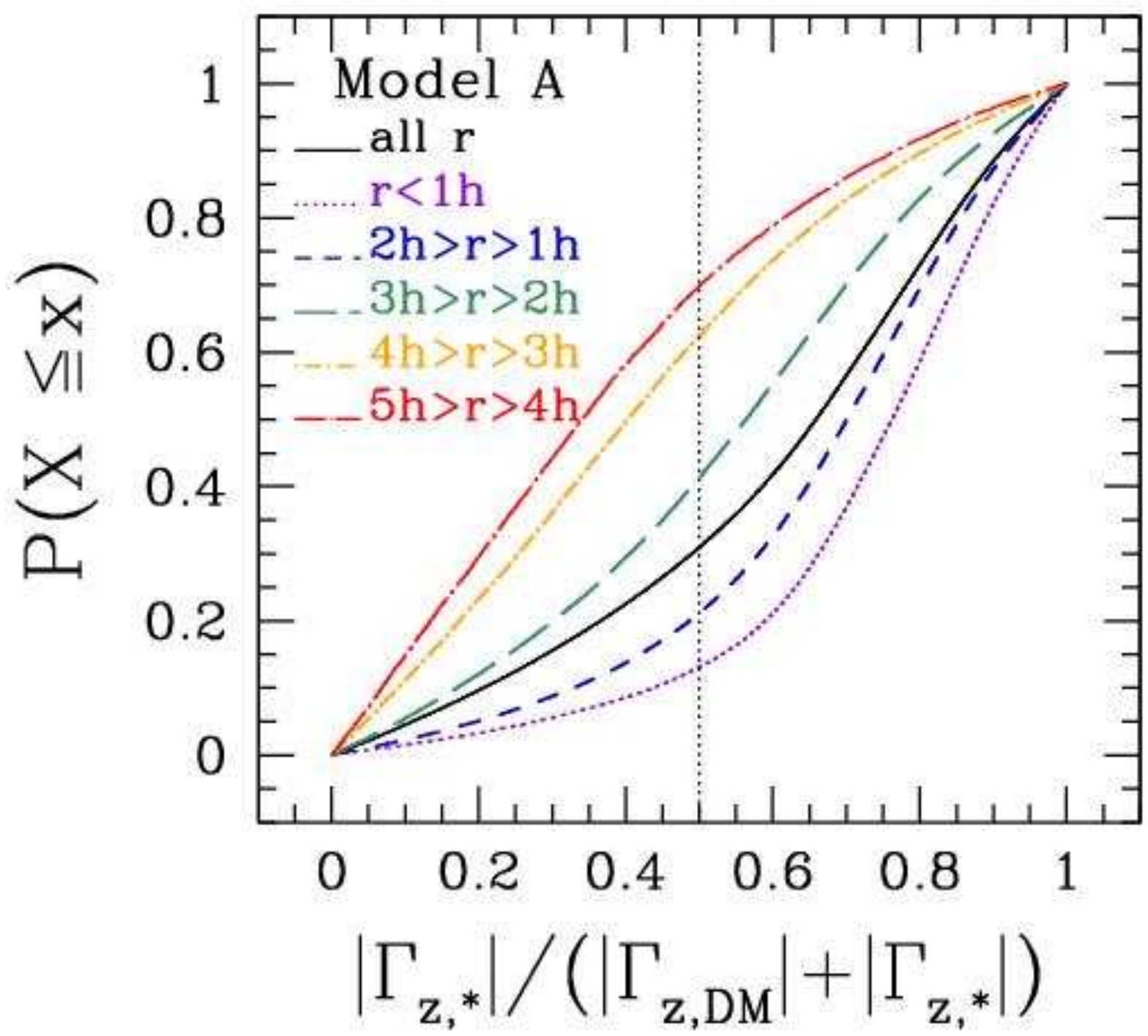}
  \includegraphics[width=80mm]{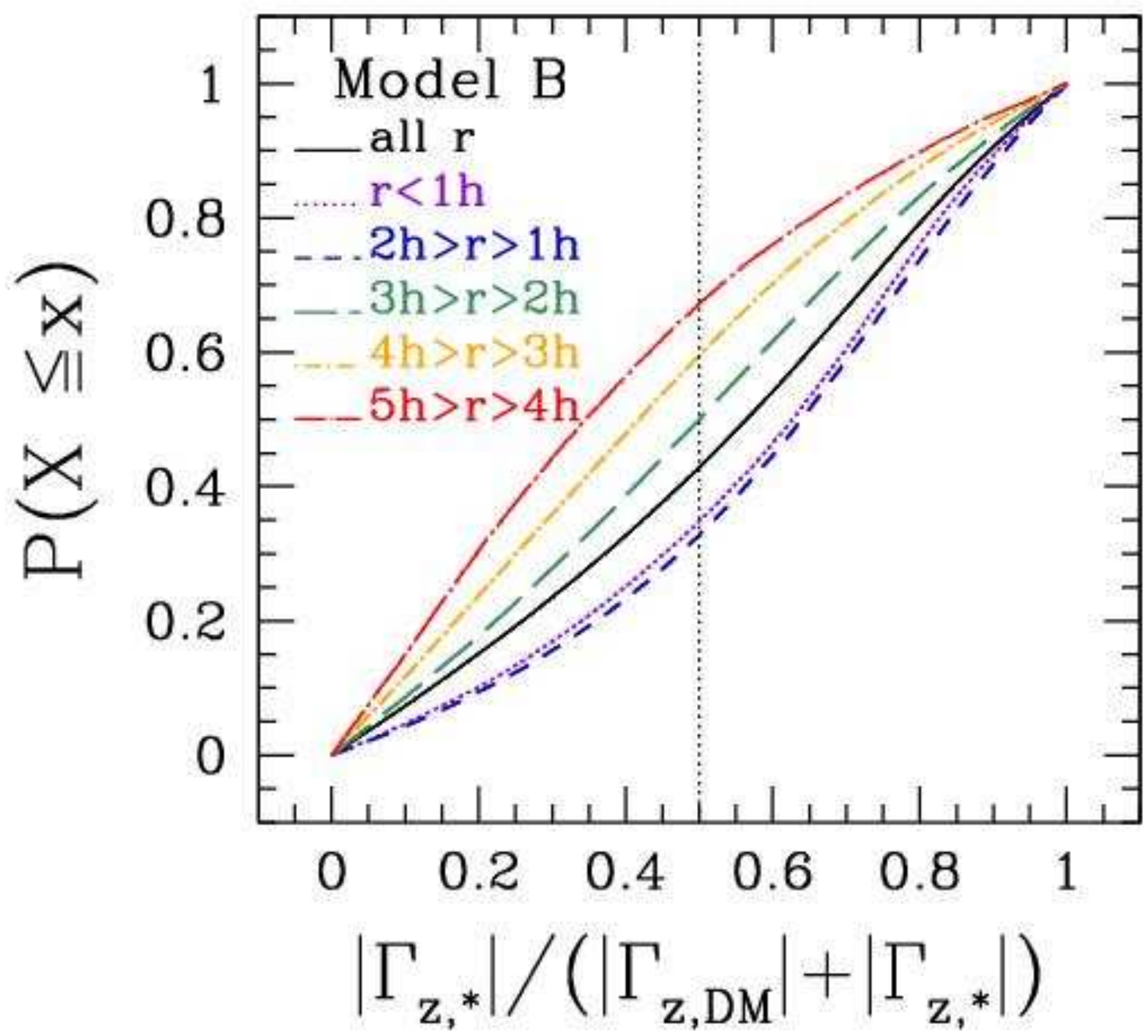}

\rm

\caption{The cumulative distribution of the ratio of the total torques on stars due to the stars compared to the total torques arising from dark matter and stars for model A and B after 2 Gyr of evolution.  $P(X \le x)$ represents the percentage of particles with less than and including the respective value represented on the $x$-axis.  The solid line represents the distribution over all radii and the colors correspond to the radial annuli listed.  This figure illustrates that within $3r_{\rmn{exp}}$ the torques due to the stellar distribution dominate those of the dark matter distribution. }

\label{cumul}

\end{figure*}



\begin{figure*}

\centering

  \includegraphics[width=80mm]{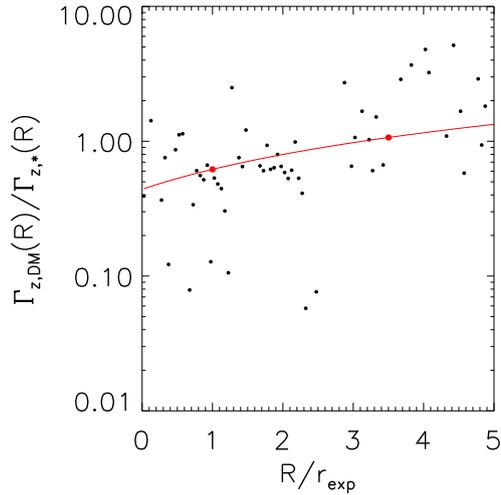}

\rm

\caption{We fit an enhancement function by choosing the median of the ratio of
  dark matter torques to the stellar torques in the inner regions ($r <2~r_{\rmn{exp}}$)
  and the outer regions ($2~r_{\rmn{exp}}<r<5~r_{\rmn{exp}}$).  Between the median points in the inner and outer regions, we plot a straight line in linear scale.}

\label{enhance}

\end{figure*}



\begin{figure*}

\centering

  \includegraphics[width=80mm]{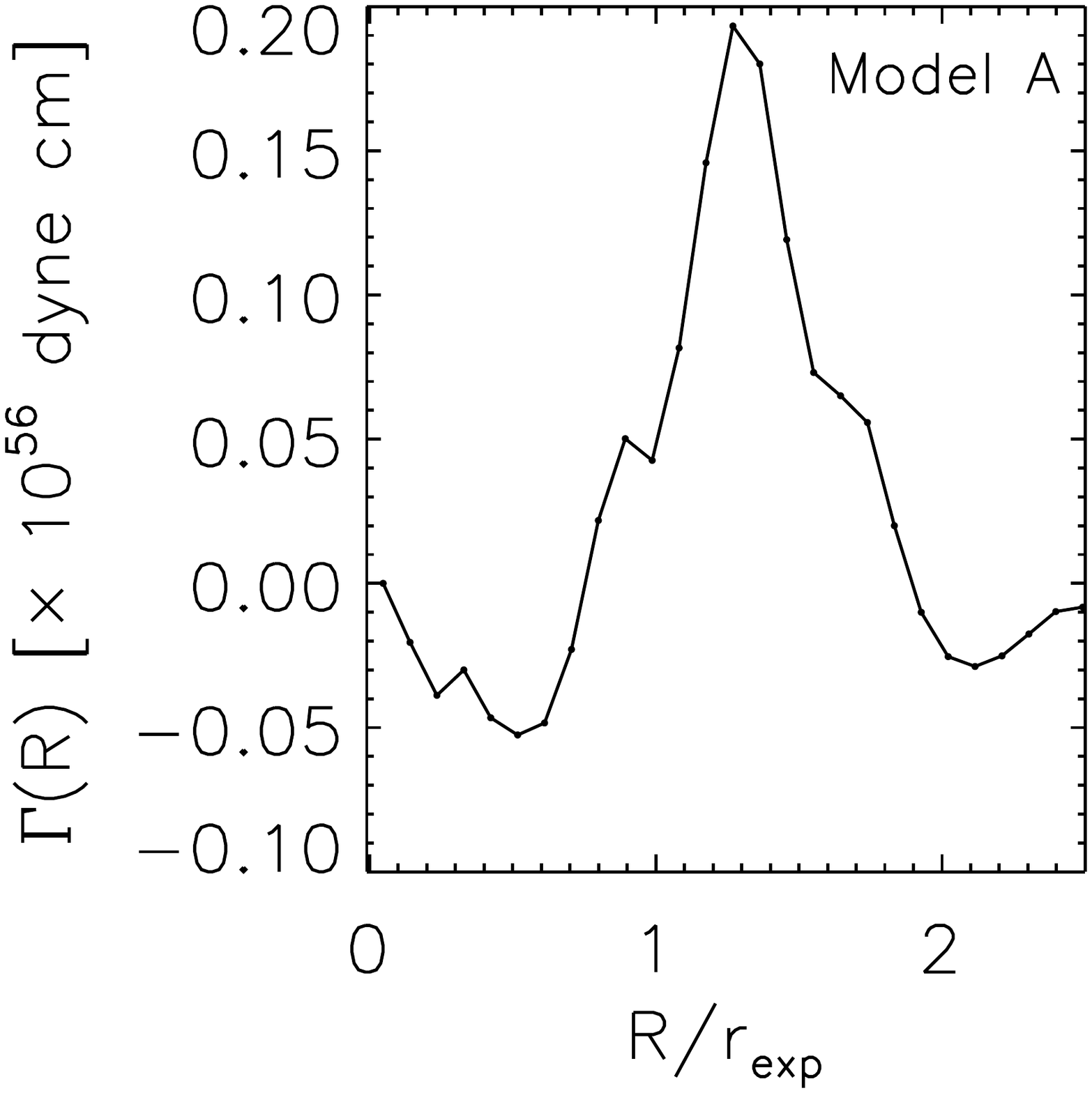}
  \includegraphics[width=80mm]{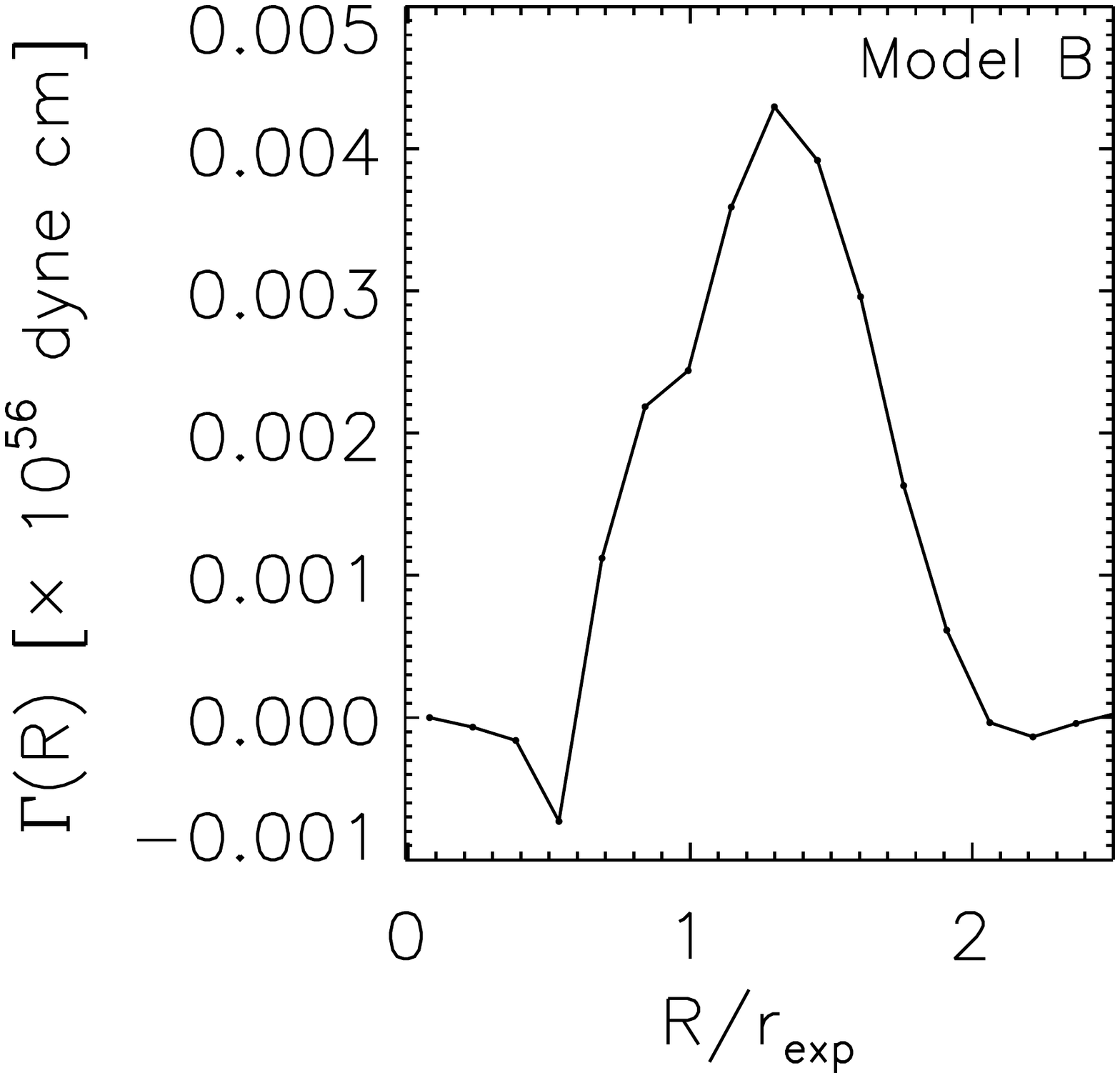}
  \includegraphics[width=80mm]{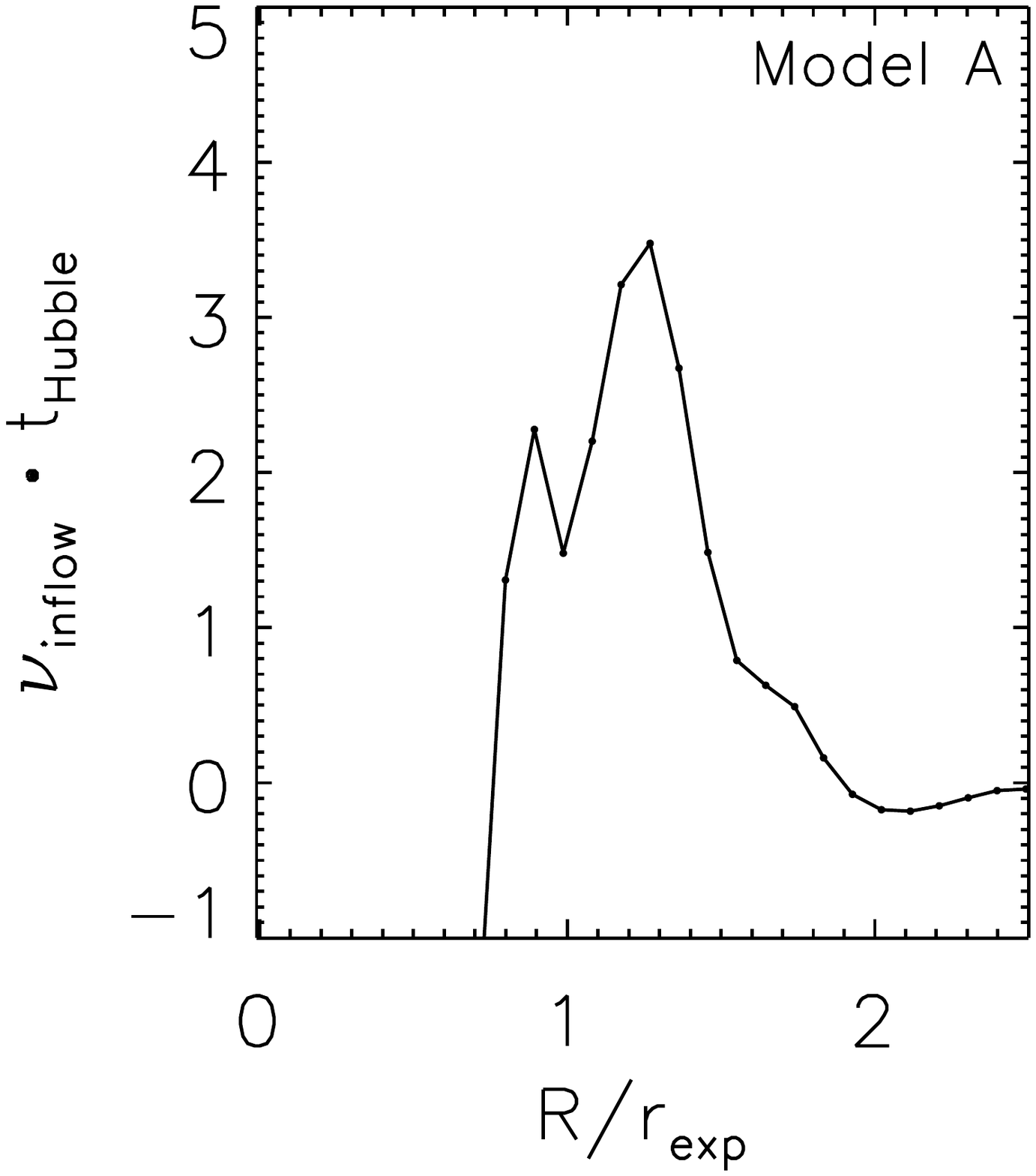}
  \includegraphics[width=80mm]{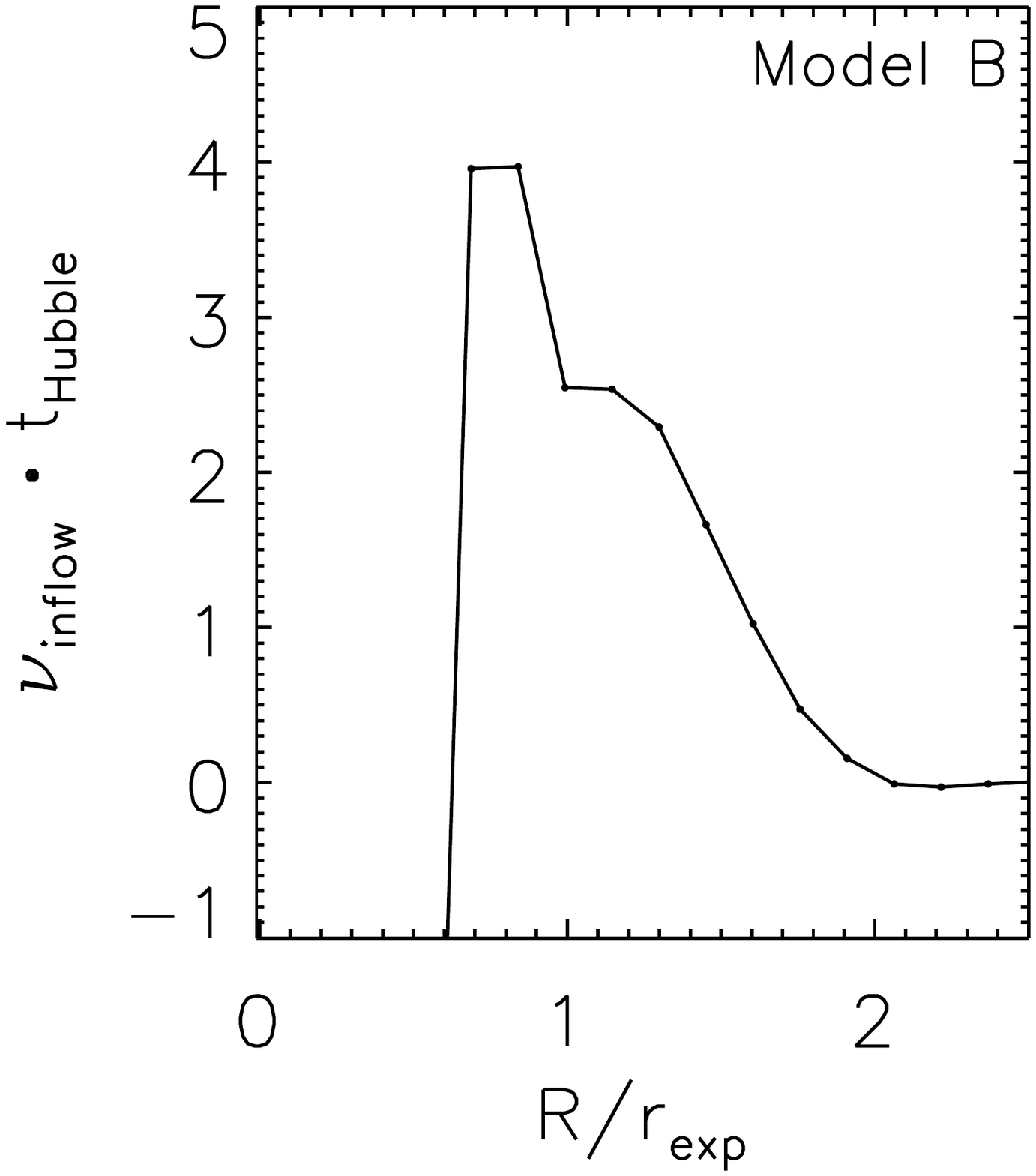}

\rm

\caption{(Upper Panel) Torque across radius R, {\it i.e.} the torque on the outer stellar
  distribution ($r > R$) due to the inner stellar distribution ($r < R$) for
  models A (left) and B (right), as described in \S 3 in following G95.  This quantity is more
  suited for application to real data; it describes the total instantaneous radial flow of angular momentum through R. For both models, the torque curve peaks between $1~r_{\rmn{exp}}$ and $2~r_{\rmn{exp}}$.  (Lower Panel) Instantaneous torque across radius $R$, assumed to reflect the instantaneous angular momentum flow, put in the secular context by presuming
  $\nu_{\rmn{inflow}} \cdot t_{\rmn{Hubble}}$ for models A (left) and B (right) where $\nu_{\rmn{inflow}}=\frac{\Gamma(R)}{J(R)}$.  We see
  that in the inner regions, $ r < 2~r_{\rmn{exp}}$, the torques imply a timescale short
  compared to $t_{\rmn{Hubble}}$ so that there is significant matter inflow within a Hubble time.  In the upper panel the amplitude of the torque curve of model A is much greater than that of model B.  Now, having scaled by the angular momentum, we see that the overall strength of the torques is comparable.  }

\label{tgnedin}

\end{figure*}



\section{The Role of Torques from Stars: a Pilot Study}

\subsection{Sample}

The simulations allowed us to test our approach, but our goal is to investigate the strength of gravitational torques in an observed sample of galaxies and use a sample average in lieu of the time integral required to determine a timescale for secular evolution.  As a pilot project, we selected 24 nearby galaxies from the 7th data release of SDSS (York et al. 2000)  .  We wanted the galaxies to be all roughly of the same mass (have the same rotational velocity) in order to use them to create a sample average.  In order to reduce errors incurred by deprojection, we also wanted them to be roughly face-on, $\cos i$ $>0.6$.  We required $m_{H} < 10$ mag with the $H$-band magnitudes taken from the 2MASS Large Galaxy Atlas (Jarrett et al. 2003).  Using the distance from NED and the Tully-Fisher relation (Tully et al., 2008), we required that the rotation velocity lie between 180 and 250 km s$^{-1}$.  The galaxies also required SDSS $g$ and $i$ band data in order to calculate the mass-to-light ratio and to make the mass maps.  Galaxies where bright stars were superimposed on the image were rejected as residuals from such features would affect the mass and torque maps.

We return to the observationally robust method of calculating the torques, $\Gamma(R)$, which we described in \S 3 (in following G95) as opposed to those employed in the simulations, $\Gamma_{\rmn{z}}$. Table 2 lists the galaxy properties in this pilot study.  In the following sections we outline how the mass maps and radial torque profiles were generated.

\begin{table*}

\caption{Sample Galaxies: Morphological classifications and distances are taken from NED.  $V_{\rmn{rot}}$ is determined via the Tully-Fisher relation and  the sign distinguishes the galaxies, which rotate clockwise (negative) and counter-clockwise (positive), assuming they are trailing spirals.  The PA, $\cos i$ and $r_{\rmn{exp}}$ were fit using GALFIT (Peng et al. 2002).  The peak value of the torque profiles $(\Gamma \cdot A)_{peak}$, where $A=\frac{Gt_{\rmn{Hubble}}}{r_{exp}^{2}V_{rot}^{3}}$ making the torques scale invariant (see \S 6) are listed for reference. }

\begin{tabular}{cccccccc}

\hline\hline

Name & Type & Distance & PA & $\cos i$ & $r_{\rmn{exp}}$ & $V_{\rmn{rot}}$ &  $(\Gamma \cdot A)_{peak}$\\

& & [Mpc] &  [deg] & $$ & [kpc] & [km s$^{-1}]$ & \\

M61 & SAB(rs)bc &17.0 & 76.0  & 0.94 & 3.0 & -222  & 0.4\\

M96 & SAB(rs)ab & 10.9 &  -15. & 0.68 & 3.0 & -234 & 1.9\\

M100 & SAB(s)bc &17.0 & 42.0  & 0.84 & 5.3 & 242 & 0.3\\

M101 & SAB(rs)cd & 4.94 & 58.0  & 0.83 & 2.5 & 201 & 0.03\\

NGC 2775 & SA(r)ab &16.7 &  -12.0 & 0.77 & 4.7 & -210 & 0.05\\

NGC 3166 & SAB(rs)0/a & 16.6 &85.0  & 0.64 & 4.0 & -259  & 0.7\\

NGC 3169 & SA(s)a  &15.1 &  59.0 & 0.71 & 4.2 & -242 & 6.4\\ 

NGC 3351 & SB(r)b  & 9.3 &  10.3 & 0.75 & 2.6 & 185 & 0.7\\

NGC 3583 & SB(s)b & 29.9 &  -57.0 & 0.71 & 3.0 & 248 & 0.5\\

NGC 3631 & SA(s)c &16.7 & -66.3 & 0.83 & 3.3 & -218 & 0.03\\

NGC 3642 & SA(r)bc & 22.9 &  -48.0  &  0.83 & 4.2 & 198 & 0.01\\

NGC 3718 & SB(s)a & 14.5 & 0.9  & 0.61 & 7.4 & -218 & 0.06\\

NGC 3938 & SA(s)c &11.6 & 34.0  & 0.93 & 2.0 & -183 & 0.02\\

NGC 4145 & SAB(rs)d & 14.3 & -78.7 & 0.59 & 3.0 & 179 & 0.04\\

NGC 4151 & SAB(rs)ab & 14.0 &  -20.0 & 0.94 & 3.6  & -229 & 0.02\\

NGC 4314 & SB(rs)a &13.2 & -62.5  & 0.84 & 2.7 & 219 & 0.9\\

NGC 4450 & SA(s)ab & 17.0 & -7.62 & 0.67 & 4.3 & -219 & 0.03\\

NGC 4579 & SAB(rs)b &17.0 &  -86.0 & 0.8 & 4.2 & 246 & 0.7\\

NGC 5317 & SA(rs)bc & 16.7 &  28.0 & 0.65 & 5.2 & -231 & 0.02\\

NGC 5383 & SB(rs)b & 32.3 & -80.6  & 0.65 & 5.7  & -252 & 0.15\\

NGC 5713 & SAB(rs)bc & 25.9 & 21.0 & 0.91 & 2.8 & 238 & 0.4\\

NGC 5921 & SB(r)bc & 20.7 &  -35.0 & 0.85 & 4.0 & 232  & 0.08\\

UGC 5001 & SB(r)0+ & 22.7 & -84.0 & 0.85 & 7.2 & 246 & 0.04\\

VCC 1253 & SB(s)0 & 17.0 &  56.6 & 0.89 & 2.67  & 204 & 0.3\\      

\hline

\end{tabular}

\end{table*}

\subsection{Stellar Mass Maps}

An accurate map of the stellar mass distribution of the galaxies is essential
to properly determine the torques.  In particular, the contrast of
non-axisymmetric features such as bars and spiral arms, must be known. Spiral
arms will contribute disproportionally to the light at any wavelength, unless
the effect is corrected.

We determine the mass distribution of the galaxies pixel-by-pixel, by
combining $g$ and $i$-band images from SDSS, as laid out in Z09.  We summarize briefly the steps here and show the importance of accurately determining the mass maps for the purposes of gravitational torque measures.

The $g-i$ colour image is used to infer stellar $i$-band mass-to-light ratios at each
pixel, which are then multiplied by the $i$-band surface brightness, resulting in the
stellar mass density.  Look-up tables as a function of $g-i$ colour  are used
to determine the $i$-band mass-to-light ratio.  These tables marginalize over
a Monte Carlo library of 50 000 stellar population models by Bruzual \& Charlot (2003, version 2007).  A median adaptive smoothing technique is used to extract
reliable flux and colour information at any position in the galaxy, while
preserving the highest possible spatial resolution.  Z09 have found that the
set of pass-bands ($g$, $i$ and $H$) works well to obtain mass maps with M/L
uncertainties $< 0.1$ dex per pixel.  Unfortunately, due to sky brightness and
long integration times in the $H$-band, it is not possible to obtain images
for a large sample of galaxies that extend beyond $2~r_{\rmn{exp}} $ using $g$, $i$ and
$H$.  Z09 have shown that combining $g$ and $i$ alone is an adequate
alternative, differing by typically $<$ 0.03 dex from mass maps made with all three pass bands.

In practice, the maps for the 24 sample galaxies are made, as follows:

\renewcommand{\labelenumi}{\roman{enumi})}

\begin{enumerate}

\item $g$ and $i$ band images of the galaxies are made from the 7th data release of SDSS (Abazajian et al., 2009, York et al., 2000) by combining the scans as our galaxies are all quite large;

\item  the sky-background was subtracted by determining the median background;

\item median adaptive smoothing with a maximum smoothing radius of 10 pixels was conducted on both images individually and $S/N > 20$ was required at each pixel (ADAPTSMOOTH, Zibetti, 2009);

\item adaptive smoothing was run again to match the smoothing in the $g$ and $i$ band;

\item $g-i$ was used to define colours for each image and a stellar mass-to-light ratio was determined via a look-up table at each pixel;

\item the mass-to-light ratio image and the $i$-band image were combined into a stellar mass map.

\end{enumerate} 

Further details of this procedure can be found in Z09.

\begin{figure*}

\centering

  \includegraphics[width=80mm]{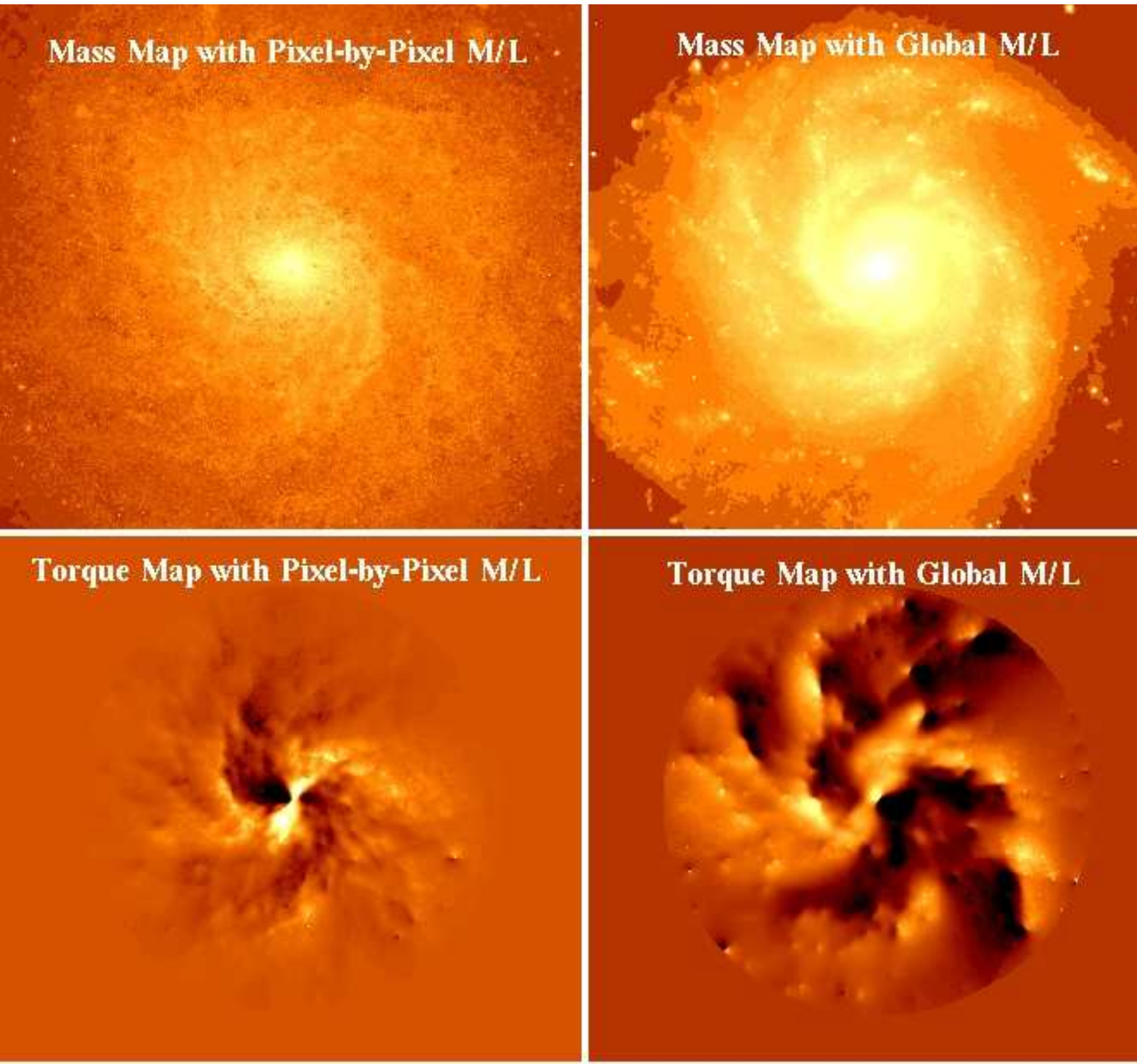}

\rm

\caption{ Mass maps (upper panels) made from the pixel-by-pixel method of Z09 using $g$ and $i$-band image from SDSS  (left) and that using a one global mass-to-light ratio {\`a} la Bell et al. (2003) using an $i$-band image for M101 (right).  The torque maps for each respective mass image are shown in panels 3 and 4 (from the left).  The global mass-to-light ratio using the Bell et al. (2003) methodology enhances the spiral arms and the torques associated with them.  We have chosen to use the pixel-by-pixel method as it avoids over-enhancing the spiral arms, which leads to torque overestimates.}

\label{M101}

\end{figure*}

It had been argued previously, that 'red' images (Elmegreen \& Elmegreen, 1985)
or near-IR images (Rix \& Zaritsky, 1995) by themselves might be close
enough approximations to the surface mass density.  Following examples from
Z09, we show this is not the case, at least for the $i$-band, in the context
of stellar torques. To illustrate this we compare the mass maps outlined here
with those generated by multiplying an $i$-band image with a global M/L.  This
global M/L was determined from the total $g-i$ colour using the constants of Bell et al.
(2003).  Fig.~\ref{M101} shows the mass and torque maps generated with these two approaches.  A comparison of the
mass maps shows that the spiral arms appear more prominent in the
$(M/L)_{global}$ mass map (the scaled $i$-band image) as opposed to the $g$
and $i$ band pixel-by-pixel mass map.  The effect of this is clearly seen in
the torque map, where strong torques are seen along the spiral arms in the
$i$-band mass map.  It is clear from the torque maps, that accurate mass maps
are central in order to correctly estimate the strengths of the torques. 

\begin{figure*}

\centering

  \includegraphics[width=80mm]{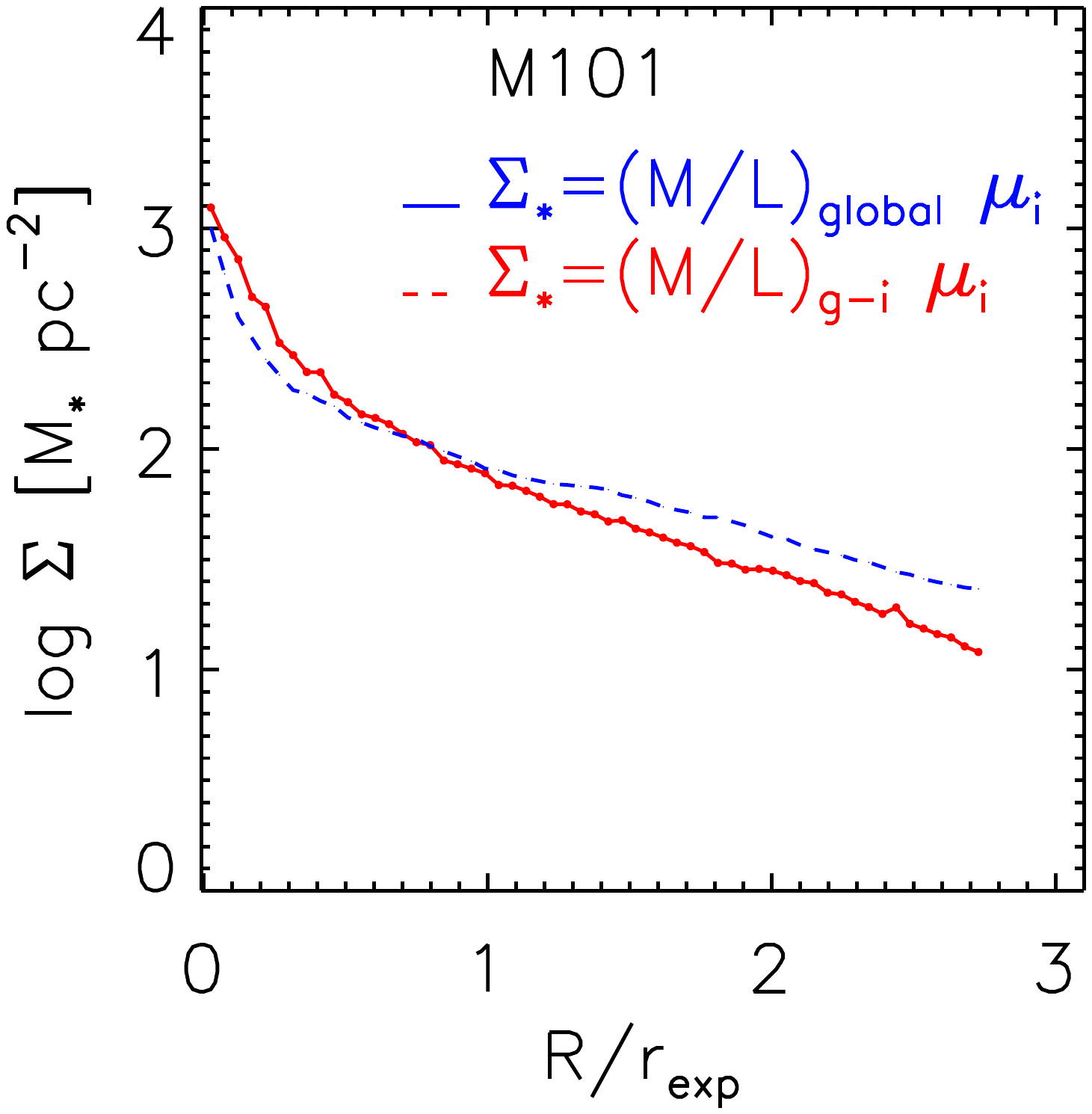}

\rm

\caption{Comparison of the mass density profiles using the mass maps made by a
  pixel-by-pixel method with $g$ and $i$-band images (blue dashed line) and that using a
  global mass-to-light ratio and the Bell et al. (2003) recipe with an $i$-band
  image (red solid line).  In the inner regions the Bell et al. (2003) recipe
  underestimates the mass, while in the outer regions it overestimates the
  mass.}

\label{M101_sb}

\end{figure*}

As pointed out by Z09, there also is a strong difference in the inferred
mass density profiles between the two approaches.   The scaled $i$-band image
(solid) in Fig.~\ref{M101_sb} tends
to underestimate the mass in the interior and overestimate the mass in the
outer parts.   The pixel-by-pixel method ensures we have correctly estimated
the mass and most importantly, mitigates the problems associated with
over-weighting regions like the spiral arms due to young, luminous
populations (Z09).


\subsection{Making Radial Torque and Angular Momentum Profiles}

After we created the mass maps using the $g$ and $i$-band images from
SDSS, we removed stars by interpolation and deprojected the galaxy in order to calculate the
torques as follows:

\renewcommand{\labelenumi}{\roman{enumi})}

\begin{enumerate}

\item We removed foreground stars using the IMEDIT task in IRAF, which replaced the stars by the average mass density in the surrounding background region;

\item The center of the galaxy was found using CENTER and represents the peak of pixel intensities in the central region;

\item We used GALFIT (Peng et al., 2002) on the $i$-band images to fit a simple exponential disk profile to the outer parts of the galaxies, in order to determine the galaxies' orientation, which included the inclination, $\cos i$, position angle, PA, and disk scale length $r_{\rmn{exp}}$.  Since the bar/bulge region can often lead to wrong fits of these values, the inner regions were masked.  The size of the masks were gradually increased until the orientation parameters converged.

\item The center values and the PA and $i$ from GALFIT were used to deproject and rotate the mass maps for all galaxies.

\end{enumerate}

The deprojected mass maps were then used to create a torque profile of the galaxy.  For each pixel we calculated the force on that pixel due to all other pixels beyond radius $R$ with an outer limit, which was set to be a maximum limit where all pixels in the annulus had sufficient signal-to-noise.  

This force included the softening correction $\epsilon=0.7r_{\rmn{exp}}/12$  in order to account for disk thickness (see \S 4.2 for further details).  The $x$ and $y$ gravitational acceleration components of the pixels beyond radius $R$, due to the mass distribution of the pixels within radius $R$, were used to determine the total $z$-component of the torque beyond radius $R$.  We express this as:
\begin{eqnarray}
\Gamma(R)=\sum_{r>R} m_{\rmn{pixel}}(x_{\rmn{pixel}}f_{\rmn{y}}-y_{\rmn{pixel}}f_{\rmn{x}}),
\end{eqnarray}
where $\sum_{r>R}$ is the sum over all pixels beyond radius $R$, $m_{\rmn{pixel}}$ is the pixel mass of the pixels beyond radius $R$, $x$ and $y$ are the pixel positions and $f_{\rmn{x}}$ and $f_{\rmn{y}}$ are the $x$ and $y$-components of the gravitational acceleration due to the mass distribution of the pixels within radius $R$ respectively.  We calculated the torque, $\Gamma(R)$, in radius increments of 5 pixels.  The galaxies have a deprojected pixel size between 60 and 120 pc.

In order to determine the angular momentum we require the rotational velocity at each pixel.  While many of the galaxies in the sample do have  kinematic measurements, we wish to outline the method we will use for a much larger sample when we will not have such information.  Thus, we construct an approximation from the image only.   The rotational velocity for each pixel was found by generating a rotation curve, 
\begin{eqnarray}
V_{\rmn{rot}}(R)=\frac{2}{\pi}V_{\rmn{max}}\arctan(\frac{R}{0.1r_{\rmn{exp}}})
\end{eqnarray}
(Courteau, 1997), where $V_{\rmn{max}}$ was found from the Tully-Fisher relation (Tully et al., 2008) and the galaxy's absolute magnitude.  The transition radius of $0.1~r_{\rmn{exp}}$ was chosen based on the mean value of 300 Sb-Sc spirals from the Courteau (1997) sample.  Given that the vast majority of spiral galaxies exhibit a trailing spiral pattern (de Vaucouleurs, 1958,  Pasha \& Smirnov, 1982 and Pasha 1985), we assume that all galaxies in our sample rotate in this way.  The sign of $V_{\rmn{rot}}$ was decided by assuming a trailing spiral structure in a right-handed system, where clockwise is negative.  Table 2 provides the rotation velocity and gives the sense of rotation.  In Appendix A we show the mass maps of the 24 galaxies and the arrows denote the direction of rotation.  For a number of galaxies, we have compared the rotation curve models to kinematic data from Sofue et al. (1999).   While the models only approximate the data, they are accurate to within 20\%.   This is sufficient for our purposes, since our results depend linearly on $V_{\rmn{rot}}$.  Note also, we are assuming that the material in the center also exhibits circular disk rotation.  While it is true that there might be considerable deviations from this in the central regions, particularly in the case of a barred galaxies where the velocities will be higher, these deviations would only make the timescales faster.  Thus, our schematic model rotation curves provide at least a minimum measure of the torque timescales.

The angular momentum inside radius $R$, $J(R)$, was found by summing the angular momentum of each pixel where $r <R$ as:
\begin{eqnarray}
J(R)=\sum_{r<R} m_{\rmn{pixel}} V_{rot,pixel}r_{\rmn{pixel}},
\end{eqnarray}
where $\sum_{r<R}$ is the sum over all pixels interior to $R$, $m_{\rmn{pixel}}$ is the mass of each pixel interior to $R$, $V_{rot,pixel}$ is the rotational velocity and $r_{\rmn{pixel}}$ is the distance to the center.

Thus, at each radial position, $R$ (in increments of 5 pixels), we have the torque, $\Gamma(R)$, on the stars beyond $R$ exerted by the stellar distribution within $R$; and we have an estimate of the total angular momentum within $R$, $J(R)$.
This defines a rate of angular momentum flow as $\nu_{\rmn{inflow}}=\Gamma(R)/J(R)$ (recall that we do not include the effects of advective transport).  In order to ascertain whether gravitational torques are strong enough to have a measurable effect within a Hubble Time, we scale this to the Hubble Time as $\nu_{\rmn{inflow}} \cdot t_{\rmn{Hubble}}$ where $t_{\rmn{Hubble}}=13.6$ Gyr.  We create radial profiles of this quantity for each galaxy in the sample. 

In order to see how sensitive the torque maps were to the steps in the image analysis, we used one galaxy to test the effects of not properly selecting the center of the galaxy and errors in the position angle, inclination and disk scale length.  We allowed the center position to vary by 4 pixels in both the $x$ and $y$ direction, which was twice the typical uncertainty in $x$ and $y$ from CENTER in IRAF.  The position angle was given a range of $\pm$ 10 degrees, the inclination angle was given a range of $\pm$ 7 degrees, and the disk scale length was given a range of 10\% (tophat range). Using Monte Carlo samplings of these error ranges, we found that  while some variation is seen in the position of the peaks of the torque profiles, the amplitude remained quite similar.  We found that the torque profiles were most sensitive to the deprojection parameters (PA and $i$) and thus we investigated this further.  

The position angle of the major axis of relatively face-on galaxies is often difficult to obtain accurately from photometry.  The challenge here is that often the faint outer regions of galaxies are not detected in surveys like SDSS, leading to photometric fits of the inner oval region.  While for a small sample of nearby galaxies it is possible to obtain kinematic measures of the deprojection parameters, for a larger sample this will not be possible.  Thus, we examined the effect of using GALFIT deprojection parameters based on photometry as opposed to kinematic estimates.  In our sample of 24 galaxies, 12 galaxies were identified with position angles measured from kinematics and in five cases the inclinations were also fit using kinematics.  Table 3 lists the GALFIT deprojection parameters (PA$_{\rmn{GALFIT}}$ and $\cos i_{\rmn{GALFIT}}$) and those from kinematics (PA$_{\rmn{literature}}$ and $\cos i_{\rmn{literature}}$).  We quantify the difference between these two sets of deprojection parameters by the angle $\alpha$, which is the angle between the kinematic and photometric deprojection vectors.  The median value of $\alpha$ is 14$^\circ$.  $\alpha$ is quite small even when there is a large discrepancy between the position angles.  This is because the galaxies are all relatively face-on and even large changes of the position angle have little effect on the overall deprojection vector, or stretch.

We show how the torque profiles compare based on the different deprojection parameters in Fig.~\ref{comp}.  The torque profiles where GALFIT deprojections were used are shown in black and those where kinematic measures were used are shown in red. The torque profiles are scaled by a factor $A=\frac{Gt_{\rmn{Hubble}}}{r_{exp}^{2}V_{rot}^{3}}$ to make them scale invariant (see \S 6).  While, the profiles do change in shape, the overall amplitude remains the same and the errors are not systematic.  Furthermore, the changes are small compared to the object-to-object variance.  The root median squared difference between the torques of the 12 galaxies based on the two sets of deprojections is 0.03.  Thus, provided the larger sample is restricted to relatively face-on galaxies, photometric deprojection parameters can be used with confidence.

\begin{table*}
\caption{Deprojection parameters of 12 galaxies with kinematic measures in the literature}

\begin{tabular}{ccccccc}
\hline\hline
Galaxy & PA$_{\rmn{GALFIT}}$  [deg] & PA$_{\rmn{literature}}$  [deg]  & $\cos i_{\rmn{GALFIT}}$ &  $\cos i_{\rmn{literature}}$ & $\alpha$ [deg] &reference\\
M61 & 76.0 & 138.0 &0.94& 0.91\footnotemark[1] & 22.8 & Cayatte et al., 1990\\
M96 & 165.0 &168.0 & 0.68 & none given & 2.3 & Schneider, 1989\\
M100 & 42.0 & 153.0 & 0.84 & 0.89\footnotemark[1]  & 48.8 & Knapen et al. ,1993\\
M101 & 58.0 & 39.0 & 0.83 & 0.95 & 17.6 & Bosma et al., 1981\\
NGC 3631 & 113.7 & 150.0 & 0.83 & 0.96\footnotemark[1]  & 22.7 & Knapen, 1997\\
NGC 3642 & 132.0	& 122.6 & 0.83 & 0.94\footnotemark[1]  & 14.5 & Verdes-Montenegro et al., 2002\\
NGC 3718 & 0.9 & 200.0 & 0.61 & 0.64 & 15.0 & Sparke et al., 2009\\
NGC 3938 & 34.0 & 20.0 & 0.93 & 0.98 & 10.8 & Van der Kruit \& Shostak, 1982\\
NGC 4450 & 172.4 &175.0 & 0.67 & 0.71\footnotemark[1] & 3.7 & Cayatte et al., 1990\\
NGC 4151 & -20.0 & 20.0 & 0.94 & 0.98 & 13.3 & Bosma et al., 1977\\
NGC 4579 & 94.0 & 95.0 & 0.80 & 0.81\footnotemark[1] & 1.1 & Cayatte et al., 1990\\
NGC 5383 & 99.4 & 85.0 & 0.65 & 0.77\footnotemark[1] & 14.3 & Sancisi et al., 1979\\
\hline
\end{tabular}
\\
${}^1$ inclination measured from photometry\\
\end{table*}

\begin{figure*}
\centering
  \includegraphics[width=100mm]{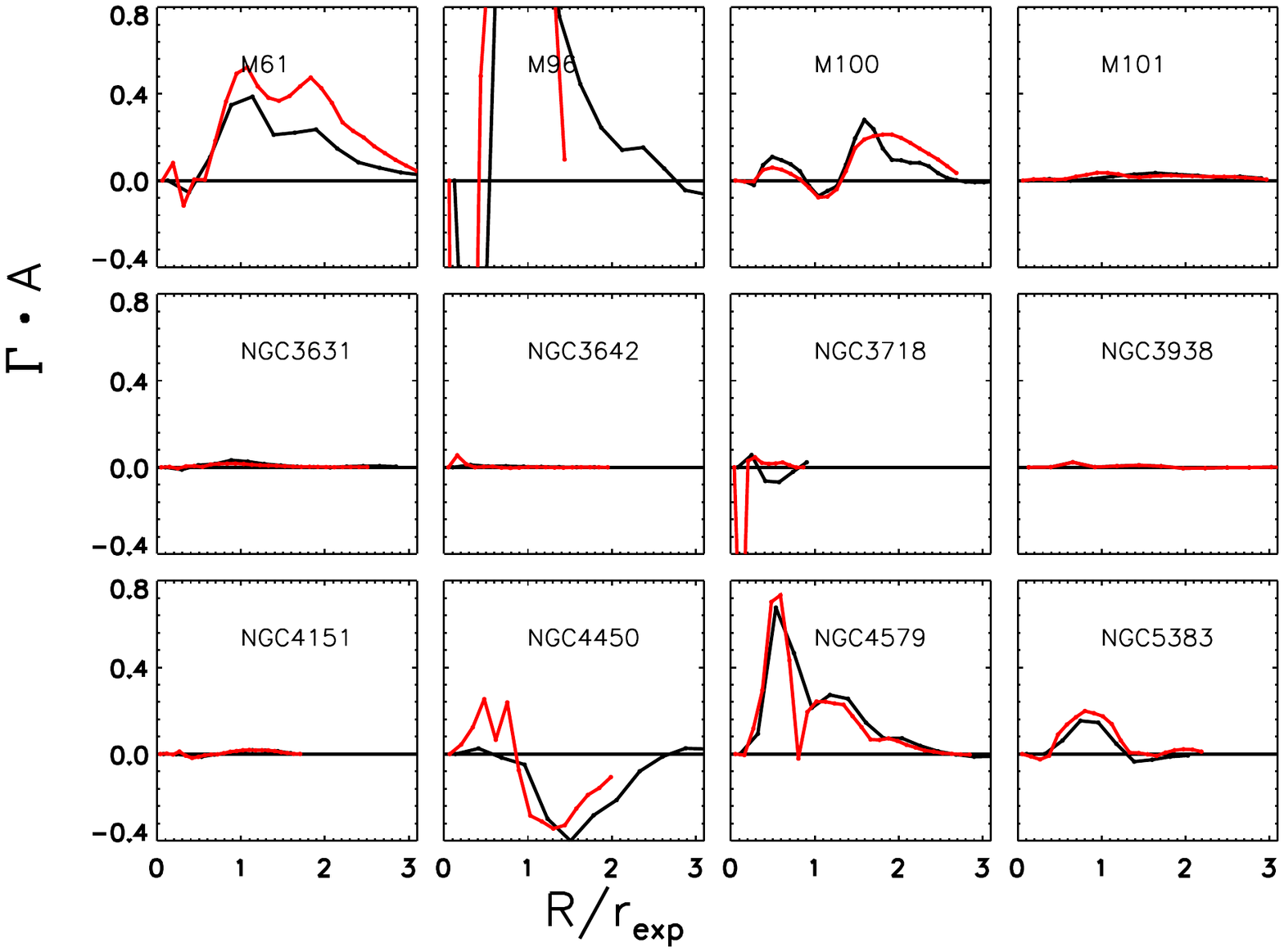}

\rm

\caption{Comparison of the effect of the deprojection values measured from GALFIT (black) to the literature values (red) on the torque profiles scaled by a factor $A=\frac{Gt_{\rmn{Hubble}}}{r_{exp}^{2}V_{rot}^{3}}$.  While the shape of the profiles change in some cases, the overall amplitude remains much the same.  The root median squared difference is 0.03.}

\label{comp}

\end{figure*}


\subsection{M100 - Comparison with the analysis Gnedin, Goodman \& Frei (1995)}

We repeated the analysis of G95 on M100 in order to compare our analysis with their results.  While we did not use Fourier decompositions, our approach was conceptually identical in that we determined the torque on the inner region of the stellar distribution due to the outer region of the stellar distribution at each radius $R$ (see Eq. 1).  G95 expressed the torques in $cm$ $dynes$ and we do so here for ease of comparison.  We used both the $i$-band image used by G95 (Frei et al. 1996) deprojected using their $PA$ and $i$ and their method to convert to mass using a constant mass-to-light ratio of 2.85 for the $i$-band as well as our mass maps made from the pixel-by-pixel method with $g$ and $i$-band images from SDSS.  In the first case we confirmed that the total stellar mass within 160\arcsec  was the same as that found by G95.  Despite this agreement, we found  substantially different results for the torque.  While our torque profile was very similar in shape (cf. their Fig. 7), the strength of the torque was found to be a factor $\sim$ 5 less than that cited by G95.  We tested our code with point masses and symmetrical distributions and confirmed that the code was indeed producing the torque amplitudes accurately.  It remains unclear as to the cause of this discrepancy.

\begin{figure*}

\centering

  \includegraphics[width=100mm]{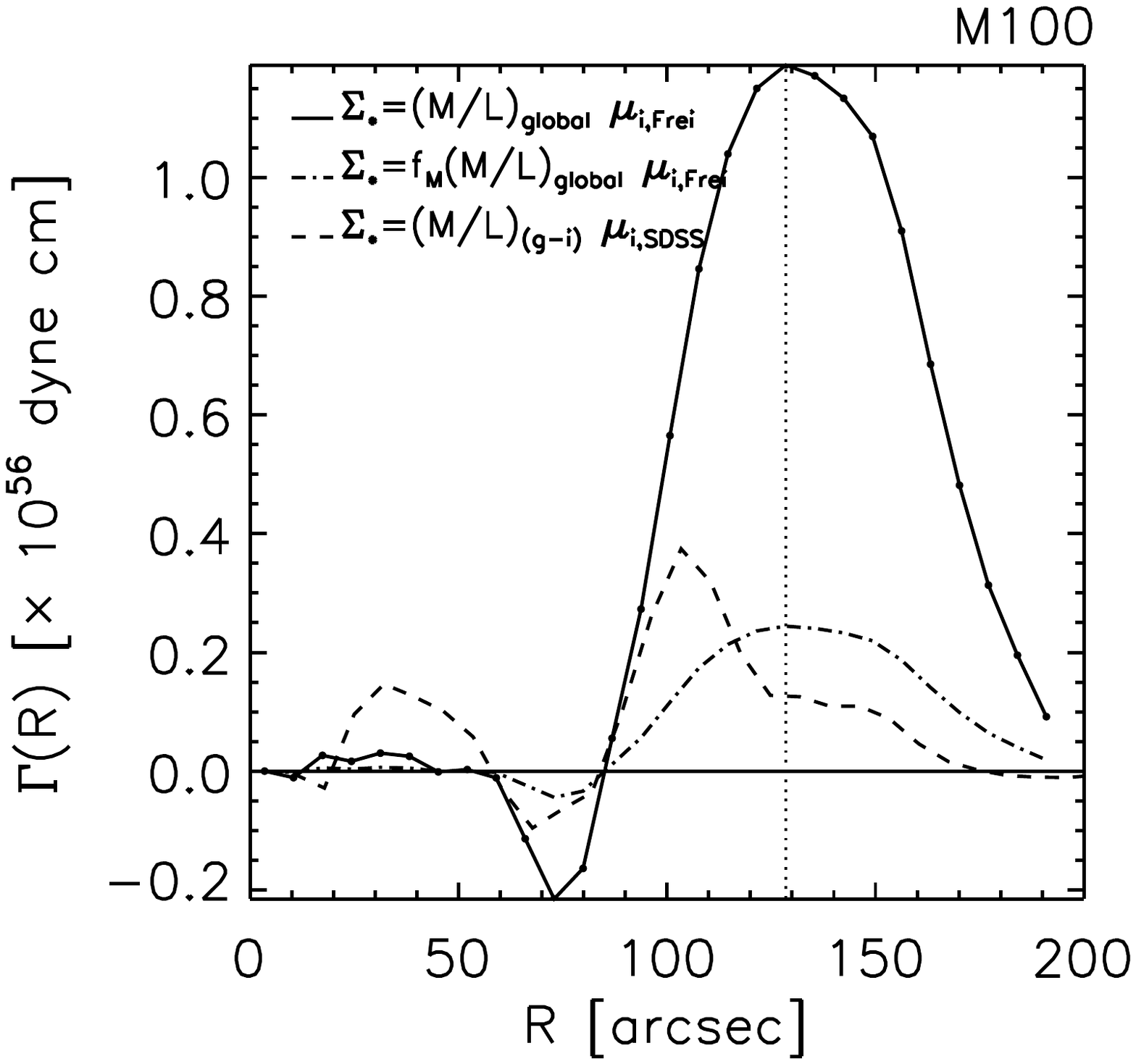}

\rm

\caption{Radial profiles of the torque for M100 using the $i$-band image of Frei et al. (1996) and a constant mass-to-light ratio (solid), the mass map made from $g$ and $i$ SDSS images using a pixel-by-pixel method (long dash) and the mass map made from the $i$-band image of Frei et al. (1996) and a constant mass-to-light ratio scaled to include the same total mass as that of the mass map made using the pixel-by-pixel method (dot-dash).   Despite repeating the same process as G95 the torques are much less than those found by them (cf. the solid line with Fig. 7 in G95).  Furthermore, there are two other issues associated with the torques calculated from the Frei et al. (1996) image.  Their total mass was a factor of 2 greater than that calculated by the pixel-to-pixel method, which means the torques are a factor of 4 greater.  We scaled the torques of the image by $f_{\rmn{M}}=0.25$ for comparison (dot-dash).    Using the more accurate pixel-by-pixel mass map, even further alters the torque profile as it avoids over-enhancing the spiral arms.  Simply using 'red' images as proxies for mass maps is insufficient in this context. }

\label{M100}

\end{figure*}

As we saw in \S 5.2, accurate mass maps are central for properly estimating the torques.  Using our $g$ and $i$-band images to create pixel-by-pixel mass maps, we found that the stellar mass was less than that assumed by G95 (by a factor of 2; see Z09  about the systematic difference with respect to earlier determinations, e.g. Bell et al., 2003).  In Fig.~\ref{M100} we show the torque curve as calculated using the image of Frei et al. (1996) and their mass-to-light ratio (solid line).  We scaled the torque curve to enclose the same total mass as that found by the pixel-by-pixel (dot-dashed).  Since the torques are proportional to the square of the mass, this implies a decrease in the torque curve by a factor of 4.  Furthermore, using the pixel-by-pixel method - as opposed to a global mass-to-light ratio - even further alters the torque profile (dashed line, \S 5.2).   This again illustrates how crucial it is to properly calculate the mass using a pixel-by-pixel method if one wants an accurate estimate of the torque.



\section{Results \& Discussion}

\begin{figure*}

\centering

  \includegraphics[width=160mm]{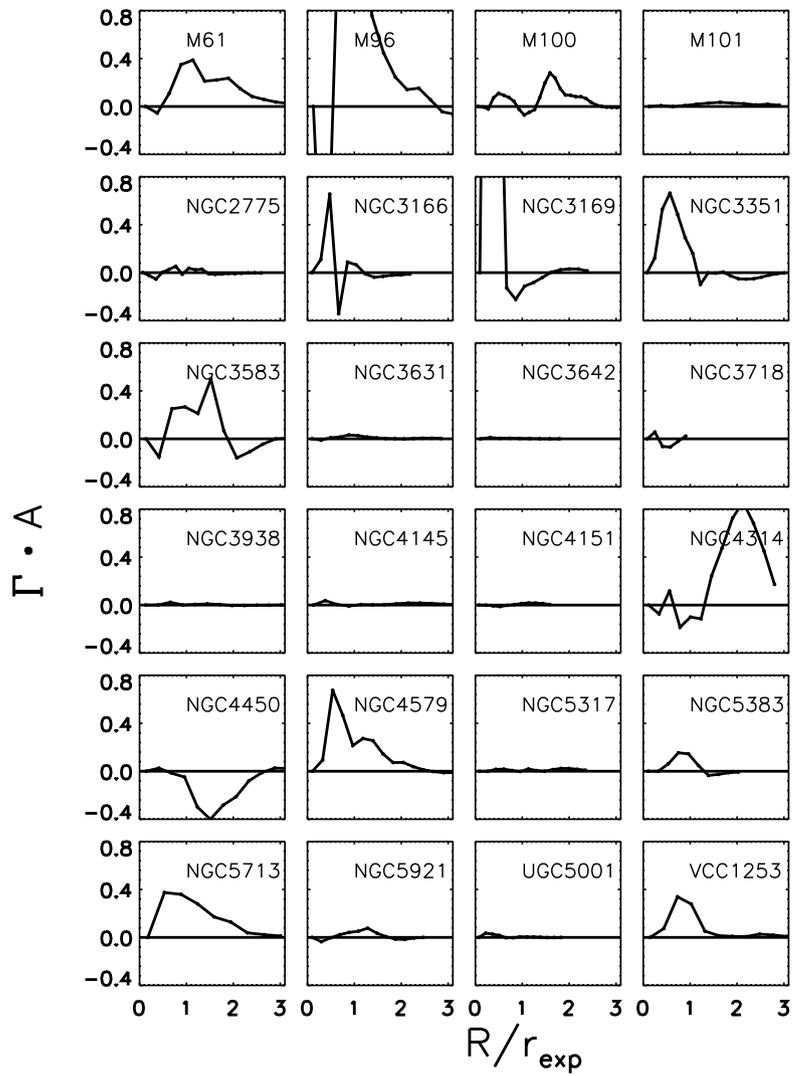}

\rm

\caption{Torque profiles scaled by $A$, where $A=\frac{Gt_{\rmn{Hubble}}}{r_{exp}^{2}V_{rot}^{3}}$.  The peak values are listed in Table 2.}

\label{panel1}

\end{figure*}

In the previous section, we have derived the instantaneous torques exerted by stars on stars across different radii, $\Gamma(R)$,  for the 24 galaxies in the sample.  For ease of comparison we scaled the torque profiles by a factor $A=\frac{Gt_{\rmn{Hubble}}}{r_{exp}^{2}V_{rot}^{3}}$, where $t_{\rmn{Hubble}}=13.6$ Gyr.  Galaxies are self-similar in terms of their disk scalelength and rotational velocity.  This scaling factor assumes the torques are self-similar and allows the torque profiles to be dimensionless. Fig.~\ref{panel1} shows all 24 profiles. While each individual galaxy shows a unique profile, a number of similarities can be identified.  The galaxies show predominately positive torques, which imply angular momentum outflow and hence matter inflow.    Thus, in much of the inner parts, angular momentum is flowing outward in these galaxies.  Furthermore, the peak of the torque curves almost all lie within $1.5~r_{\rmn{exp}}$.   This agrees with our simulations, where we also found the peak of the torque curve to lie within $1.5~r_{\rmn{exp}}$.  Thus, both the simulations and observations show that, instantaneously, angular momentum is flowing outward in the inner disk and the peak value lies at approximately one disk scale length.  As we have seen in \S 3, how gravitational torques lead to matter flow is complex, but at least qualitatively angular momentum flow is related to matter flow.

n Fig.~\ref{panel1} we see that a third of the galaxies display strong torques in the inner regions (the peak values of these curves are listed in Table 2).  By strong, we mean torques that have inflow timescales shorter than a Hubble Time.   One may note as well that those galaxies with strong torques also show strong bars.  There are two possible interpretations here; either only barred galaxies have strong torques and these galaxies have bars, which are permanent features that do not dissolve over time or we can we interpret the relative abundance of galaxies with strong torques as an estimate of their duty cycle.  In the case of the latter, we see that most of the time the galaxies are in a state such that the torques are not significant, however at some times the torques are very strong.  In order to quantify the length of time in periods of strong and weak torques, one would need to turn to a much larger sample.  Given that barred and unbarred galaxies have similar dynamical and structural parameters (Courteau et al. 2003) it seems sensible to treat them as a single group.  However, since the dissolution of bars remains a topic of debate (e.g. Berentzen et al.1998, Bournaud \& Combes 2002, Bournaud et al. 2005, Debattista et al. 2006, Curir et al. 2008) we intend to examine sub-samples of barred and non-barred galaxies with a larger sample.  If it is the case that a fraction of galaxies are always barred, while others, with similar global properties, never or rarely are, it will be interesting to examine whether secular changes due to the bars, leave that sub-population indistinguishable in their gross profiles.
 
We note, that M100, in contradiction to what G95 found, does not exhibit strong torques such that they would be significant within a Hubble Time.  As we saw in \S 5.4, when we repeated the analysis of G95, our calculations show significantly weaker torques and longer timescales than theirs.  G95 found a timescale of $\sim$ 5-10 Gyr, while we find a timescale of $\sim$ 25-30 Gyr, which is significantly longer than a Hubble Time.  This implies that angular momentum redistribution is not important for M100 at this point in time.

While it is interesting to look at the individual galaxy profiles, they only reflect instantaneous torque curves and one cannot assume that the torques of the past and future will be the same, since the spiral and bar features may change over time.  Thus, in  order to draw conclusions on the overall strength of the gravitational torques in spiral galaxies, we must turn to an ensemble average, as a proxy for the observationally infeasible time integral required to derive a mean rate of secular evolution.  Given that direct observations (Jogee et al. 2005) suggest that the characteristic strength of spiral arms and bars has not changed in the last 5-8 Gyrs, or, if it has, the effect has been moderate (by a factor of 2), we can assume for an ensemble that $\langle \Gamma(R,t_{\rmn{now}}\rangle_{\rmn{sample}} \approx \langle \Gamma(R,t_{\rmn{later}})\rangle_{\rmn{sample}}$.  In this approximation, sample stacking becomes a substitute for time-averaging.  In order to average the galaxies, we scaled them first by disk scale length.  Alternative methods, which we explored, involved scaling by bar length or the distance of the torque profile peak, $r_{\rmn{peak}}$.  Bar length is known to be proportional to disk scale length (P\'erez et al. 2005) and may thus be equivalent.  However, in the case of unbarred galaxies, it is not clear how to scale them.  Scaling by $r_{\rmn{peak}}$ will by definition create a peak in the averaged torque plot.  We aimed to have an independent means to scale the galaxies and have thus selected disk scale length.

\begin{figure*}

\centering

  \includegraphics[width=80mm]{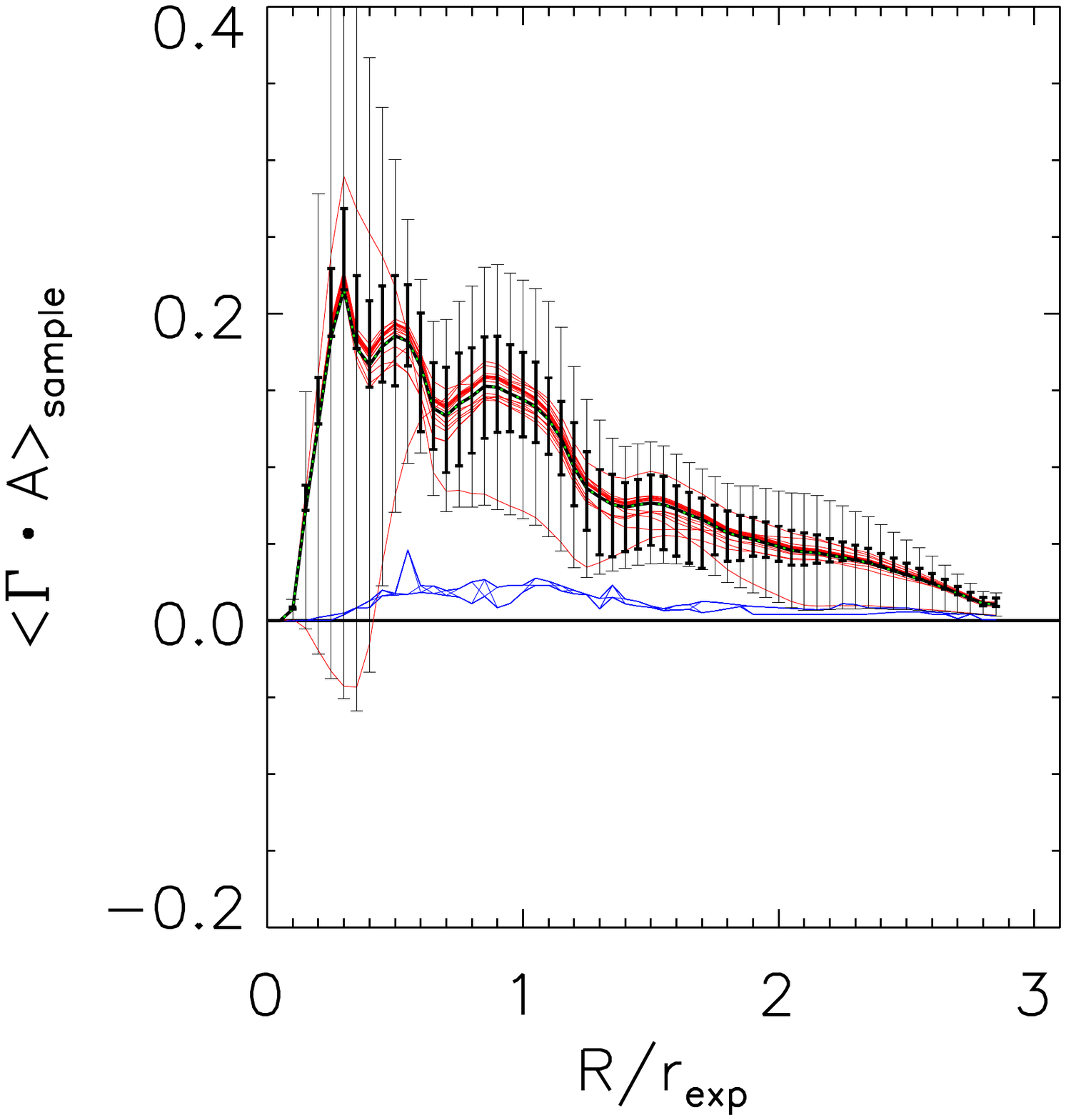}
  \includegraphics[width=80mm]{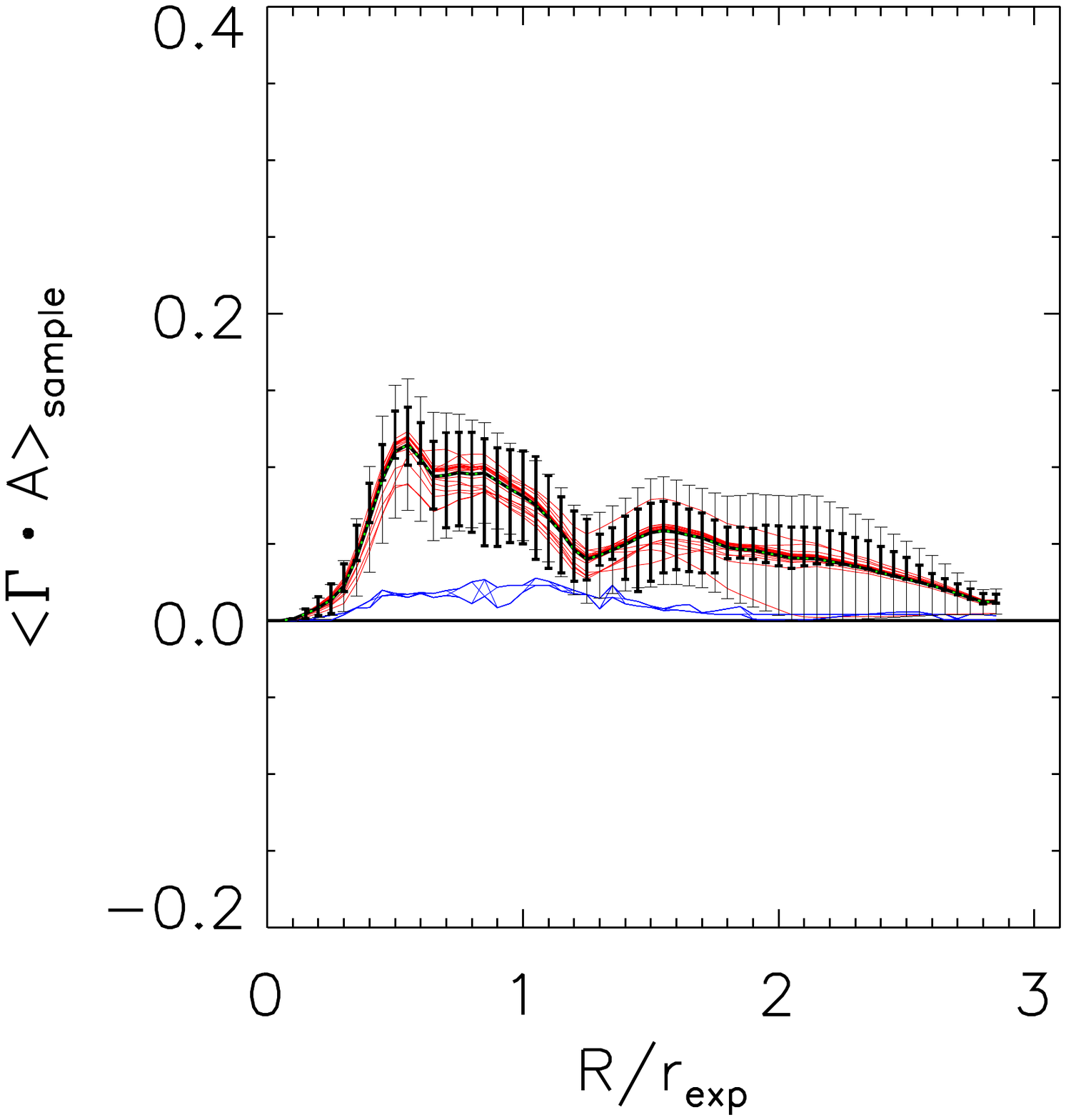}

\rm

\caption{Mean torque profiles for the sample of 24 galaxies calculated (left) and for 22 galaxies (right).  The red curves denote the jackknife iterations of the mean and the blue curves denote the jackknife iterations of the median.  The long error bars show the jackknife $\sigma$ and the short thick error bars show the 68\% confidence limit.  We see that the torque peaks in the inner part of the disk and that there is consistent angular momentum outflow and matter inflow within $3~r_{\rmn{exp}}$.}

\label{bootpeak}

\end{figure*}

Fig.~\ref{bootpeak} shows the results of averaging the torque profiles over the sample of 24 galaxies, after scaling  the torques by $A=\frac{Gt_{\rmn{Hubble}}}{r_{exp}^{2}V_{rot}^{3}}$.    For the purposes of a time average, it is essential to use the mean instead of the median of the sample.  The median provides the most common state a galaxy is in, whereas the mean provides an average over time.  The qualification of the uncertainties (as indicated by the error bars) bears careful consideration. We have used the jackknife method to determine the uncertainties on the mean torque profile. The red curves show the jackknife iterations for the mean of the sample and the blue curves show the jackknife iterations for the median.  The solid black line shows the mean of the entire sample. The long error bars in Fig.~\ref{bootpeak} show the jackknife standard deviation of the mean, $\sigma_{J,mean}$ which is calculated as
\begin{eqnarray}
\sigma_{J,mean}(R)=\sqrt{\frac{N-1}{N} \sum_{i=1}^{N} (x_{J,i}(R)- \bar{x}(R))^2},
\end{eqnarray}
where N is the number of galaxies, $x_{J,i}(R)$ is the mean torque for each jackknife iteration at the position $R$ and $\bar{x}(R)$ is the mean torque for the entire sample.  The short error bars replace $\sigma$ by the 68 \% confidence range multiplied by $\sqrt{N-1}$.  If the distribution were gaussian, we would expect that these two sets of error bars would be the same.  This implies that the statistic is affected by a small number of objects, which may reflect rare stages of galaxy evolution or more mundane data issues. 

The blue curves which represent 24 median jackknife iterations show substantially lower torques than the red curves, which represent the mean.  This implies that much of the time the galaxies are in a state where the torques are relatively weak, but for some periods of the time, the torques are much stronger.  In order to determine how long the torques are strong, one would need a much large sample.  

Fig.~\ref{bootpeak} shows that there is a characteristic profile for the angular momentum flow within $3~r_{\rmn{exp}}$: the torques are positive implying that angular momentum is transported outward.  Previous works (Kranz et al. 2003 and Tamburro et al. 2008) have shown that the co-rotation pattern of spiral arms lies at $2.7~r_{\rmn{exp}}$.   Taken together, this constitutes direct observational proof that for disk galaxies as a class angular momentum is indeed transported outward, as expected from theoretical studies.  This observational result is independent of whether spiral arms are quasi-stationary as we have used the sample average.  

If one examines closely the 24 red mean curves of the jackknife iterations one notes that two curves deviate significantly from the mean of the entire sample.  This implies that two galaxies are contributing strongly to the mean torque curve.  These galaxies are M96 and NGC 3169.  It is possible that these galaxies represent time spans in an average galaxy's life where the torques are significantly different and thus they should be kept in the sample.  However, they may also be peculiar cases.  NGC 3169 had a bright foreground star superimposed, which proved challenging to subtract.  There is concern that poor subtraction of this star may have caused anomalies in the final mass map.  Furthermore, NGC 3169 is an interacting galaxy that has clearly been disturbed.  Due to the large bar in M96, it was difficult to converge on values for the position angle and inclination during the GALFIT iterations.  Given these issues, we show in the right panel the sample average calculated by removing these two galaxies.  In order to determine if the torque profiles of these galaxies are truly representative of periods of strongly deviating torques, one requires a much larger sample.

\begin{figure*}

\centering

  \includegraphics[width=80mm]{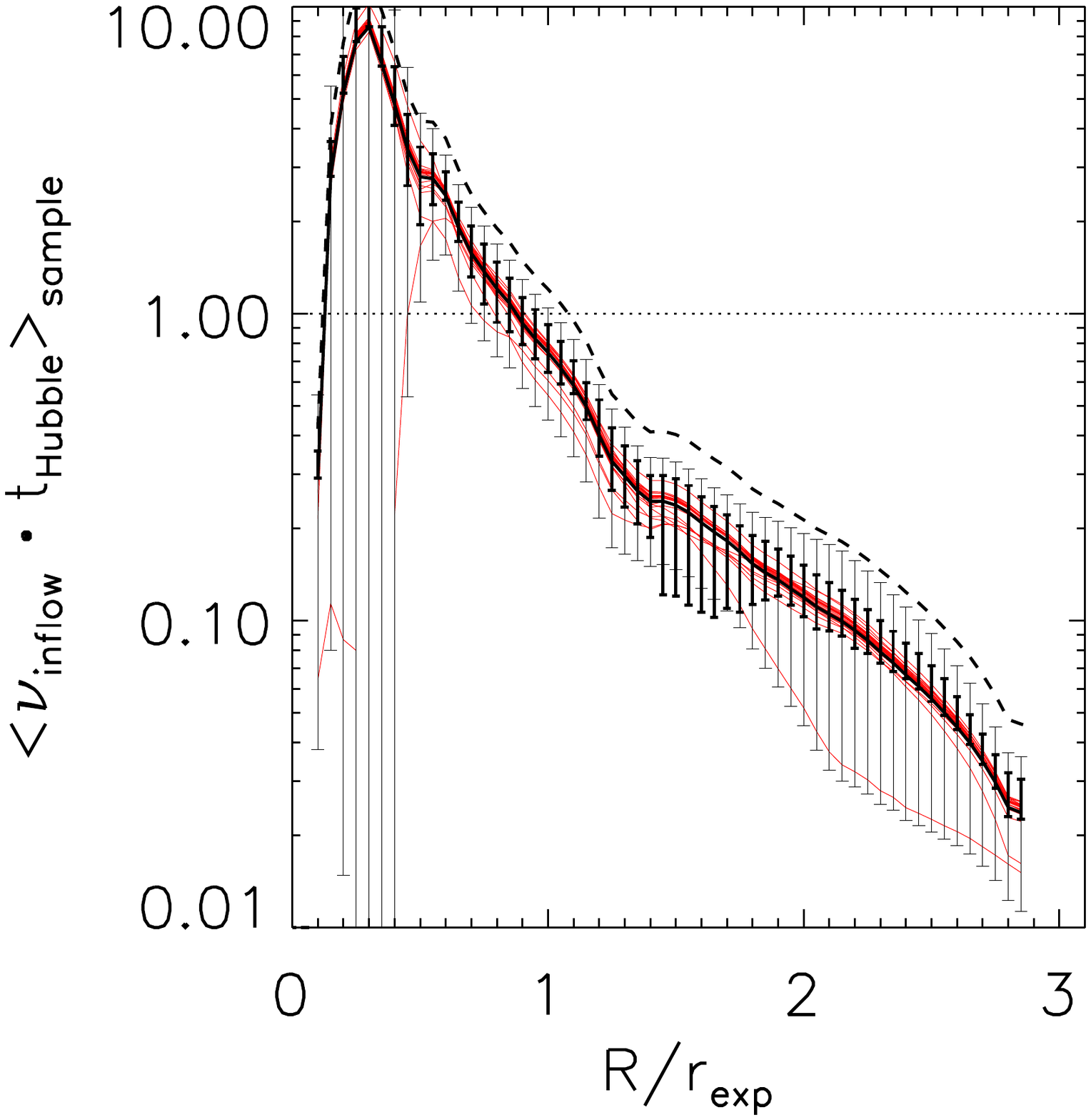}
 \includegraphics[width=80mm]{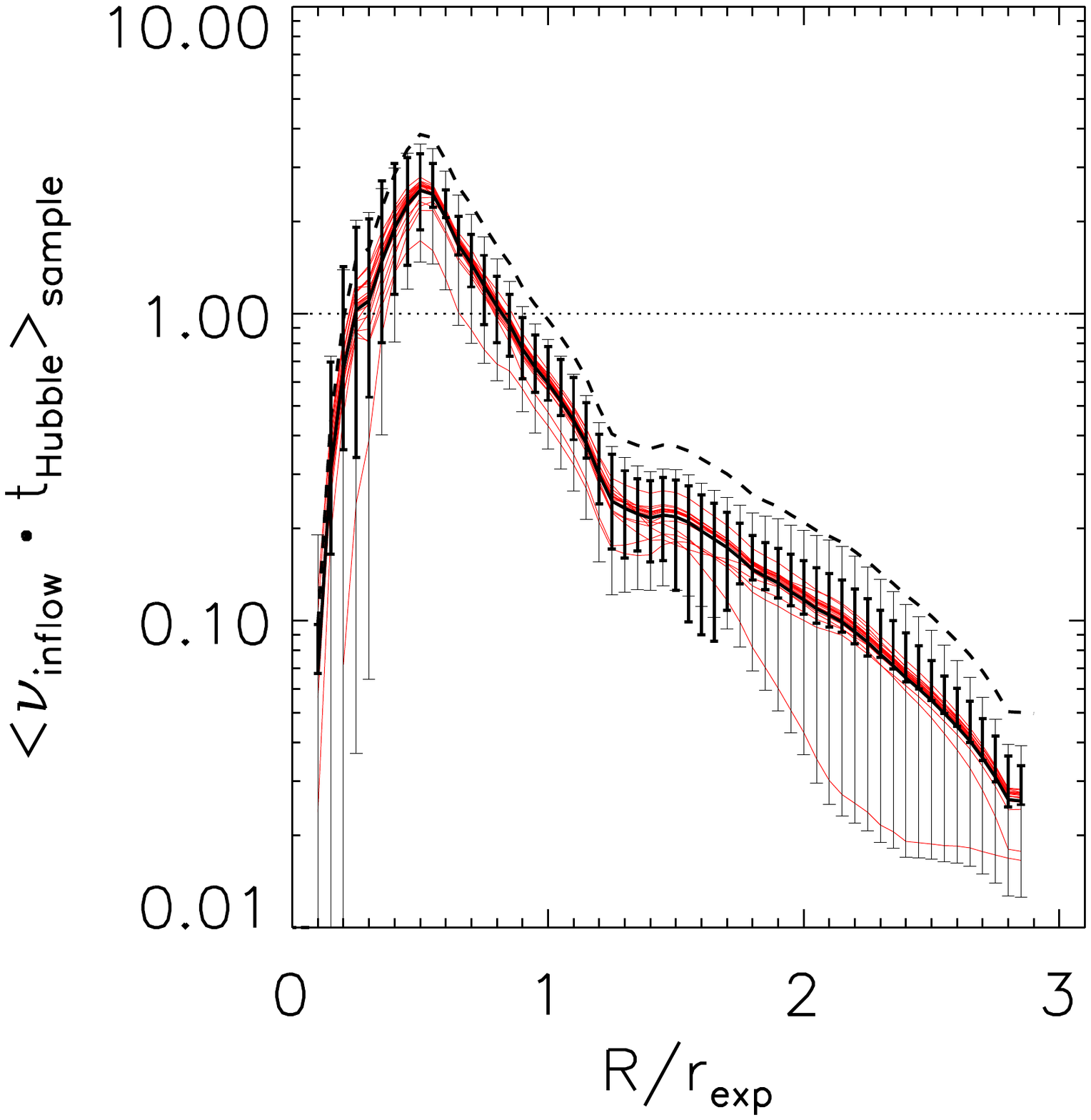}

\rm

\caption{Mean $\nu_{\rmn{inflow}} \cdot t_{\rmn{Hubble}}$ profile for the sample of 24 galaxies (left) and for 22 galaxies (right).    The red curves denote the jackknife iterations of the mean.  The long error bars show the jackknife $\sigma$ and the short thick error bars show the 68\% confidence limit. We see that within $1~r_{\rmn{exp}}$ the timescale for matter inflow is less than a Hubble Time.  The dashed line shows the enhancement in the inner regions due to the dark matter torques, determined from the simulations in \S 4.}

\label{bootflow}

\end{figure*}

Fig.~\ref{bootflow} quantifies the rate of angular momentum transport.  Again we show the 24 jackknife iterations (red curves) and the mean curve of all 24 (black).  On the right we have removed M96 and NGC 3169 from the sample, leaving 22 galaxies.  For both cases in the very inner regions ($r < 1~r_{\rmn{exp}}$), the torques on the stars due to stellar distribution lead to a flow rate which is less than a Hubble Time.  We remind the reader that we have simplified this picture and that  gravitational torques do not trivially translate into matter flow  (see \S 3 for a discussion of this).   It is plausible, however, that the corresponding matter inflow is such that the radial profile will change over time.  In the inner regions, the timescale for secular evolution is $\sim$ 4 Gyr.   The timescale depends linearly on $V_{\rmn{rot}}$ and, as explained in \S 5, our model rotation curves differ in some cases by 20\% to those found in the literature, which would lead to a 20\% enhancement or reduction of the timescale. 

The timescale is based solely on the torques due to the stellar distribution.   In \S 4 the simulations showed that the dark matter distribution can also exert torques in the same direction as the stellar distribution.  In the inner regions the torques may be enhanced by as much as 20\% and in the regions beyond $2~r_{\rmn{exp}}$, the enhancement may be as much as 50\%.  The dashed line in  Fig. ~\ref{bootflow} shows the effect of the dark matter enhancement using the enhancement factor derived from a fit of $\frac{\Gamma_{\rmn{z,DM}}}{\Gamma_{\rmn{z,*}}}$ in \S 4.   Beyond $1~r_{\rmn{exp}}$, the rate of angular momentum flow becomes considerably longer than a Hubble Time (Fig.~\ref{bootflow}).  Thus, we expect that the radial profile will remain as formed beyond this position.  

Given that the stellar profile in the inner regions may change, as stars are funneled inward, our findings suggest that such a process may be effective in creating so-called pseudo-bulges ({\it e.g.} Kormendy \& Kennicutt, 2004), while the outer regions change little.  In the simulations of Foyle et al. (2008) it was found that if a two-component profile develops in a galaxy, the outer disk scale length will remain true to the initial value, while the inner disk scale length will evolve over time to lower values as matter funnels inward to the bar and bulge.  These observations substantiate the findings of their simulations.  Simulations of Valenzuela \& Klypin (2003) also showed that the matter exchange occurred solely in the very central regions of the galaxies (within 2 kpc).




\section{Conclusion}
Previous studies that have addressed the significance of secular evolution due to gravitational torques have taken either a computational or theoretical approach ({\it i.e.} Lynden-Bell \& Kalnajs, 1972, Bertin, 1983, Sellwood \& Binney, 2002, etc.). We present here a direct observation-based estimate for the long-term torque-driven angular momentum transport .  We have extended and improved upon the work of G95, by using a larger sample of 24 galaxies and constructing accurate mass maps using $g$ and $i$-band images from SDSS and a pixel-by-pixel mass-to-light ratio (Z09).  These mass maps allowed us to derive instantaneous torque maps based on the stellar mass distribution and compute radial matter inflow profiles.  We stacked the 24 inflow profiles scaled by disk scale length in order to use an ensemble average to bypass the time integral required to investigate the long-term strength of secular evolution.

We tested the validity of this approach using N-body/SPH simulations, where we verified that the torques due to the stellar distribution are stronger than those from the dark matter distribution.  Thus, torques derived only from the stellar distribution are a conservative (lower) bound for the total torque and we derived a correction for the torques due to the dark matter distribution.  We also verified that the uncertainties in deprojecting the galaxies, finding the centroid and determining the disk scale length were moderate.

For each galaxy in the sample we derived instantaneous torque profiles scaled by disk scale length.  We translated these into angular momentum flow profiles.  However, we did not include the effects of advective transport, since there is no way to measure these effects observationally.  For the individual galaxies we found that there was consistently angular momentum outflow in the inner regions and the peak of each torque curve was within $1.5~r_{\rmn{exp}}$.  The ensemble properties of the stacked sample of 24 galaxies show that for typical present day spiral galaxies:
\renewcommand{\labelenumi}{\roman{enumi})}
\begin{enumerate}
\item angular momentum is transported outward within $3~r_{\rmn{exp}}$;
\item within $1~r_{\rmn{exp}}$ the torques lead to an angular momentum outflow timescale much less than a Hubble Time;
\item  the average angular momentum flow timescale in this inner region is $\sim$ 4 Gyr;
\item beyond $1~r_{\rmn{exp}}$ the timescale is much longer than a Hubble Time, implying that the stellar radial profile has largely remained unchanged.
\end{enumerate} 
In conducting this study we have found that accurate mass maps are central in properly determining the torques due to the stellar distribution.  Global mass-to-light ratios ({\it e.g.} scaled $i$-band images) are not sufficient and pixel-by-pixel mass-to-light ratios ensure that the total mass is properly estimated and that regions with young, blue, luminous populations, like spiral arms are not over-weighted (Z09).

Our results show that, for spiral galaxies as a class, secular evolution is {\it observed} to be important within one disk scale length and may lead to the formation of pseudo-bulges.  Beyond one disk scale length however, the timescales are such that little to no change in the radial profiles should be expected. We stress that our simple picture must be qualified since advective transport and the fact that angular momentum flow does not directly translate into matter flow has not been treated.

This has been a preliminary pilot study and we are extending this project to a much larger sample of galaxies.  We will make comparisons between different spiral types and barred and unbarred galaxies in subsequent papers.

\section*{Acknowledgements}
Kelly Foyle acknowledges generous support from the Max Planck Society, International Max Planck Research School for Astronomy and Cosmic Physics at the University of Heidelberg and the National Science and Engineering Research Council of Canada.  We acknowledge stimulating discussions with E. Athanassoula, B. Elmegreen, S. Courteau and E. Bell.  We acknowledge correspondences with O. Gnedin and S. Jogee.  We wish to thank the referee, A. Bosma, for his thorough reviews of this paper and many helpful ideas and suggestions.  We also wish to thank J. Sellwood for his review and clarification of the underlying physics at play.


\clearpage
\onecolumn


\appendix

\section{Mass Maps for Sample}

We show the final mass maps after deprojection for the sample of 24 galaxies.   The arrows denote the direction of rotation chosen based on the assumption of a trailing spiral.


\begin{figure*}

\centering

  \includegraphics[width=100mm]{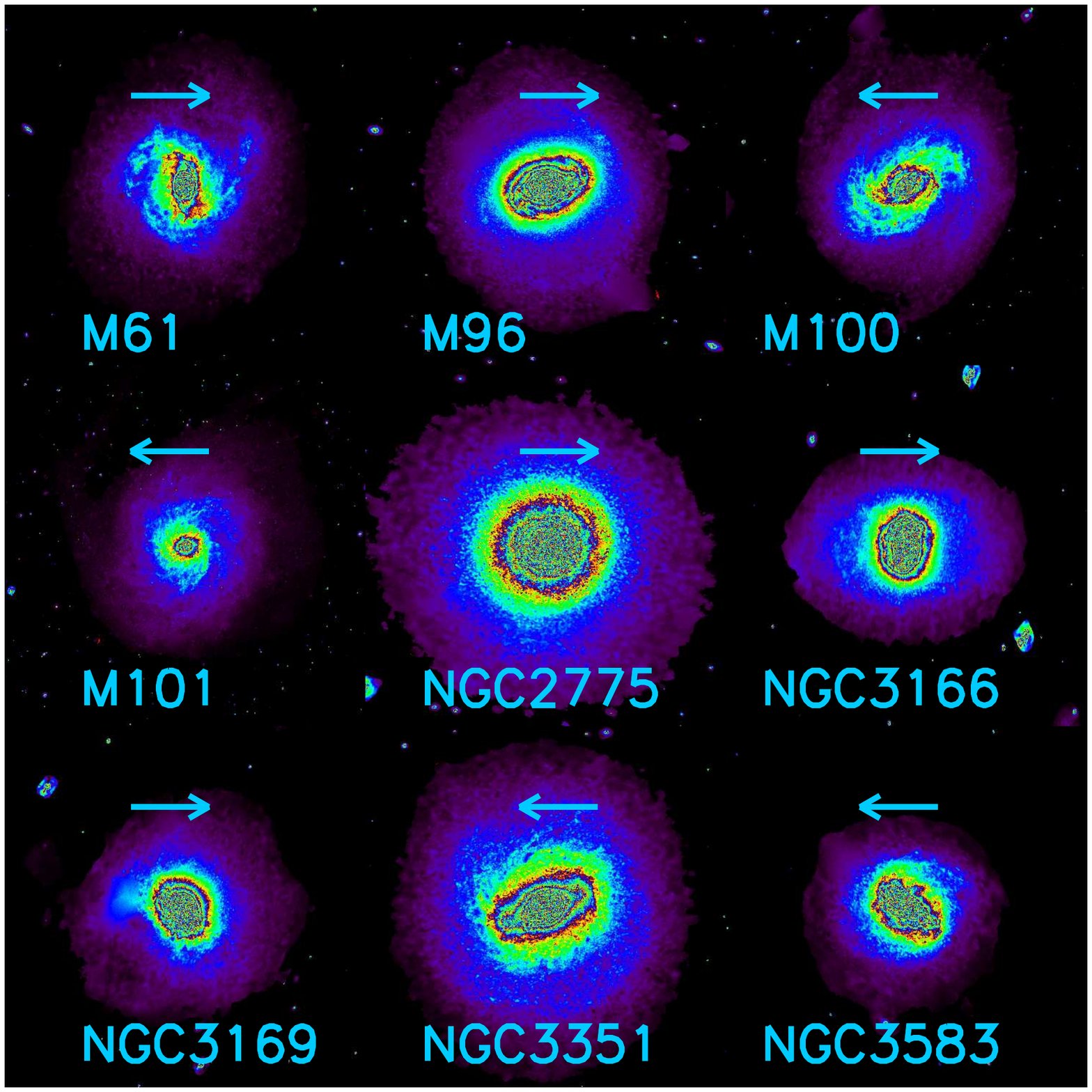}

\rm

\caption{Mass maps and direction of rotation for M61, M96, M100, M101, NGC 2775, NGC 3166, NGC 3169, NGC 3351 and NGC 3583.}

\label{m4}

\end{figure*}



\begin{figure*}

\centering

  \includegraphics[width=100mm]{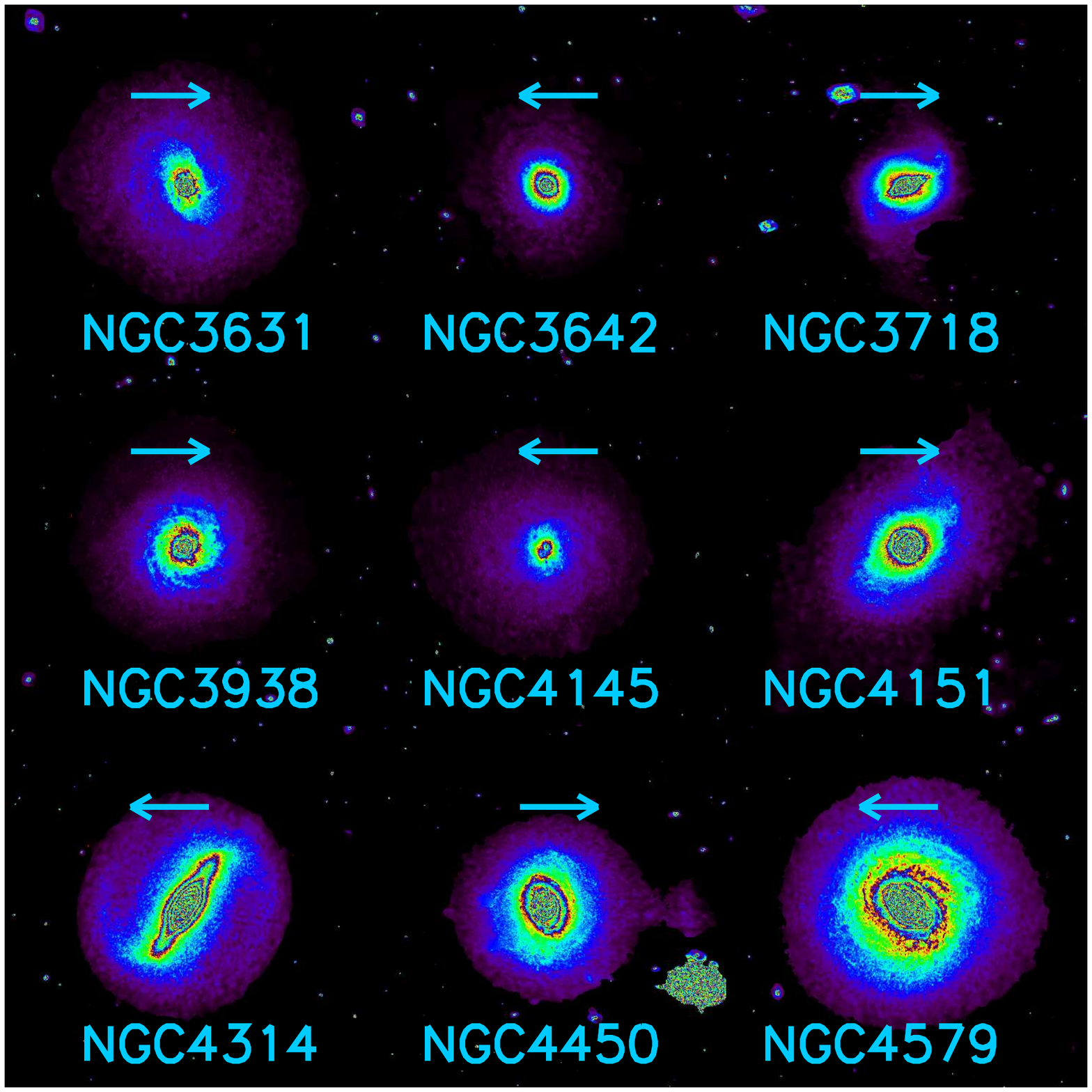}

\rm

\caption{Mass maps and direction of rotation for NGC 3631, NGC 3642, NGC 3718, NGC 3938, NGC 4145, NGC 4151, NGC 4314, NGC 4450 and NGC 4579.}

\label{ngc1}

\end{figure*}



\begin{figure*}

\centering

  \includegraphics[width=100mm]{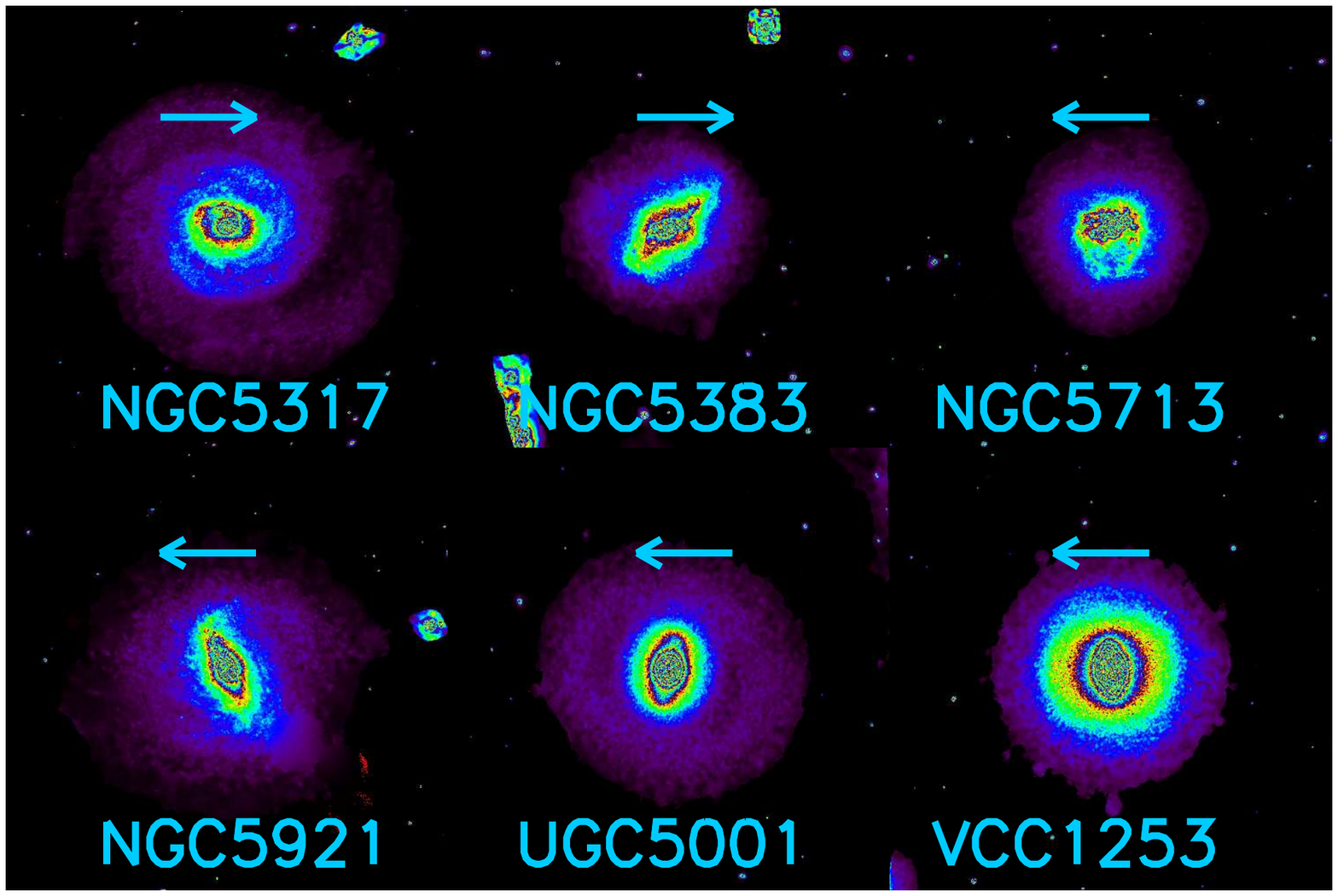}

\rm

\caption{Mass maps and direction of rotation for NGC 5317, NGC 5383, NGC 5713, NGC 5921, UGC 5001 and VCC 1253.}

\label{ngc2}

\end{figure*}


\label{lastpage}

\end{document}